\documentclass[12pt]{article}
\usepackage{amsmath,amsthm, amssymb}
\usepackage{graphicx}
\usepackage{leqno}
\usepackage{hyperref}
\numberwithin{equation}{section}
\begin{document}
\title{\vskip-40pt  Interaction-Induced Wave Function Collapse Respects Conservation Laws}
\author{Edward J. Gillis\footnote{email: gillise@provide.net}}

\maketitle

\begin{abstract} 

\noindent

Because quantum measurements have probabilistic outcomes they can seem to violate conservation laws in individual experiments. Despite these appearances, strict conservation of momentum, orbital angular momentum, and energy can be shown to be consistent with the assumption that the entangling interactions that constitute measurements induce a real collapse of the wave function. The essential idea is that measured systems always have some pre-existing entanglement relations with (usually larger) systems, and that apparent changes in conserved quantities in the measured system are correlated with compensating changes in these larger systems. Since wave function collapse is mediated by entanglement relations a full accounting of the relevant quantities requires a computation over all interacting, entangled systems. The demonstrations by Gemmer and Mahler[1], and by Durt[2,3], that entanglement is a generic result of interaction are central to the argument. A stochastic collapse equation based on interaction potentials is described and shown to guarantee conservation of the relevant quantities at all stages of evolution.

\end{abstract}

\section{Introduction}
\label{intro}

Symmetries insure that the total momentum, angular momentum, and energy of a closed physical system are strictly conserved under unitary evolution. But when measurements intervene, the seemingly nonunitary changes in the state of the system can appear to violate the standard conservation laws in individual experiments. The range of responses to this situation has included denials that the apparently nonunitary changes are genuine physical occurrences, assertions that the uncertainty principle renders the violations innocuous, and claims that the conservation laws hold only statistically and that the relevant quantities are conserved when experimental results are averaged over a large enough set of similar cases.

This note presents a reevaluation of the status of the conservation laws based on the assumption that the apparent collapse of the  wave function associated with measurement is a real, nonlocal occurrence that is induced by the kinds of entangling interactions that constitute measurements. I argue that when all relevant entanglement relations are taken into account, it can be seen that the conservation laws are respected in individual cases. The argument turns on considering both the entanglement relations of the measured system generated during the measurement process and those resulting from prior  interactions. It relies heavily on the demonstrations by Gemmer and Mahler\cite{Gemmer_Mahler} and by Durt\cite{Durt_a,Durt_1} that entanglement is a generic result of interactions between systems. These demonstrations have important implications for how one should frame the question of whether quantities such as momentum, angular momentum, and energy are conserved in measurement processes. Given the web of entanglement relations that must be taken into account in these situations, they imply that the ``total system" over which the relevant quantities are computed must include all subsystems that have interacted with the measured (sub)system in a significant way, \textit{including any preparation apparatus}.

The need to treat all of the systems involved as quantum systems, and to take all relevant interactions into account is often obscured by the simplified descriptions of idealized experiments. The idealizations that are used in pursuit of mathematical simplicity can lead one to overlook the way in which conserved quantities are shared among subsystems. To properly assess the effects of measurement on these quantities we must allow for the possibility that they are shared, and define them in a way that is consistent with  the general theorem relating conservation laws to the symmetries of the Hamiltonian. That theorem states that a quantity, $\mathbf{q}$, is conserved if the operator, $\mathbf{Q}$, with which it is associated commutes with the Hamiltonian: 
\begin{equation}\label{1p1}
(d/dt)\langle\,\psi|\hat{\mathbf{Q}}|\psi\,\rangle \; = \; 
\frac{i}{\hbar}\langle\,\psi|\hat{\mathbf{H}}\hat{\mathbf{Q}} - \hat{\mathbf{Q}}\hat{\mathbf{H}}|\psi\,\rangle \; = \; 0.  \end{equation} 
Thus, the conservation laws tell us that the quantities of interest are properly defined by the expression: 
\begin{equation}\label{1p2}   
\mathbf{q} \; = \;\langle\psi|\hat{\mathbf{Q}}|\psi\rangle,
\end{equation}  where $\psi$ 
includes all entangled subsystems (both microscopic and macroscopic). The term usually applied to quantities defined in this way is ``expectation value", but this terminology can be very misleading. It results in the tendency to regard $\mathbf{q}$ as being well-defined only when the subsystem being measured happens to be in an eigenstate of the observable, $\mathbf{Q}$.

That such a restrictive definition is completely inadequate to deal with questions about the proper understanding of conservation laws should be obvious when one considers the reasons that we find these laws interesting and useful. Aside from the somewhat trivial cases in which an isolated system propagates freely, conservation laws tell us what happens to various quantities when these quantities are exchanged during interactions between subsystems.

Conserved quantities are almost always shared among different subsystems. This sharing is very obvious when we consider systems that involve only a small number of elementary subsystems. For example, we have no trouble in seeing that a singlet state of two spin-one-half particles has total (spin) angular momentum of zero. But the possibility that a conserved quantity can be shared is almost completely overlooked when we consider the relationship between microscopic and macroscopic systems. If quantum theory is the correct description of nature then, ultimately, all (sub)systems have to be treated as quantum systems, and entanglement relations between them have to be taken into account. With this perspective the question of conservation becomes whether $\mathbf{q}_{final}$ can be shown to be arbitrarily close to $\mathbf{q}_{initial}$ by considering all relevant systems and taking into account enough history. I will argue that conservation does, in fact, hold in this sense.

The assumption that wave function collapse is a real, nonlocal effect, and that it is induced by measurement-like interactions is crucial to the argument developed here. It is important to emphasize that collapse involves a network of entangled subsystems - not just an isolated elementary system in a product state. It is mediated by entanglement relations, and when it occurs it collapses the entire wave function to one of its branches. The branches are defined by the correlations of the quantities that have been exchanged in the interactions that generated the entanglement. When branches are either eliminated or enhanced by the collapse the effects of prior exchanges are also eliminated or enhanced. The nonlocal change of state thus results in a physical record of exchanges that is completely consistent with the conservation of the relevant quantities.

By focusing on the interactions that are responsible for both the exchange of conserved quantities and the generation of entanglement relations we pick out the basis into which the wave function collapses in the most natural way. This focus also leads, very naturally, to a stochastic collapse equation. The collapse operator is defined in terms of interaction potentials with a time dependence related to the rate at which entanglement is generated. By examining the evolution of the wave function described by this equation in configuration space, it can be shown that momentum and orbital angular momentum are strictly conserved at every stage of evolution.

The stochastic collapse equation also conserves energy within the limited range of accuracy provided by a nonrelativistic formulation. It insures strict conservation in all situations in which the Schr\"{o}dinger equation \textit{correctly} predicts energy conservation. In addition, and most importantly, it shows that the apparent increase in kinetic energy associated with the narrowing of wave packets during collapse is correlated with compensating changes in the systems with which it has become entangled through prior interactions.

The very small discrepancies with strict conservation implied by the collapse equation occur in exactly the same situations in which the predictions of the standard nonrelativistic  Schr\"{o}dinger equation become less accurate, and they are of the same magnitude as the inaccuracies. The only forms of energy that the conventional theory can describe are potential energy and nonrelativistic kinetic energy. The small deviations in kinetic energy due to relativistic mass increase (of order $ \approx 10^{-4} \, - \, 10^{-5}$ at low speeds) are not accounted for, nor is the  radiation associated with accelerated motion and changes of state. These discrepancies should be seen as indications of the limitations of a nonrelativistic account, rather than as actual breaches of the conservation law. These issues are addressed in considerable detail in Section 7 and Appendix B.

The structure of the argument is as follows. Section 2 briefly reviews the way in which interaction generates entanglement and assesses the effect this has on the definition of conserved quantities. Section 3 illustrates the main argument of the paper with a simple example. It examines the often hidden entanglement that results from an interaction between a microscopic and a macroscopic system (such as a preparation apparatus) in order to show that this very small amount of entanglement is sufficient to insure that conservation laws are respected in individual cases when measurements induce the collapse of the wave function. Section 4 describes a stochastic generalization of the Schr\"{o}dinger equation based on interaction potentials. Section 5 attempts to determine what are the most reasonable initial conditions to assume in examining the issue of conservation. Section 6 provides a proof of strict conservation of momentum and orbital angular momentum. Section 7 addresses the issue of conservation of energy. Section 8 presents some additional illustrations of the general argument. Section 9 summarizes the discussion.

 \section{Interaction, Entanglement, and Conserved Quantities}
 \label{sec:2}

Whenever two systems interact they have some effect on one another. For quantum systems the interaction leads to some entanglement between them as demonstrated by Gemmer and Mahler\cite{Gemmer_Mahler} and by Durt\cite{Durt_a,Durt_1}. One can see this by examining what happens during the interaction between two previously unentangled systems. Represent the combined system as $\psi(\vec{\mathbf{w_1}},\vec{\mathbf{w_2}})$, and the component subsystems as $\phi(\vec{\mathbf{w_1}})$ and $\chi(\vec{\mathbf{w_2}})$. If the subsystems did not interact then $\psi(\vec{\mathbf{w_1}},\vec{\mathbf{w_2}})$ would remain a simple product of the two components: 
$ \psi(\vec{\mathbf{w_1}},\vec{\mathbf{w_2}}) \; = \; 
\phi(\vec{\mathbf{w_1}})\chi(\vec{\mathbf{w_2}}). \, $
Because kinetic energy operators act on the subsystems separately their wave functions would evolve independently. To represent the interaction between the systems one must introduce a potential,  $\mathbf{\hat{V}}(\vec{\mathbf{w_1}},\vec{\mathbf{w_2}})$, with a nonzero gradient, $  \mathbf{\nabla} \mathbf{\hat{V}}(\vec{\mathbf{w_1}},\vec{\mathbf{w_2}}) \neq 0. 
\, $ The interaction alters the simple-product structure of the state as we can see by examining  the potential energy term in the Schr\"{o}dinger equation, 
\begin{equation}\label{2p1}
 d\psi({\vec{\mathbf{w_1}},\vec{\mathbf{w_2}}})\, / dt  \,   
= \, (-i/\hbar) \mathbf{\hat{V}}(\vec{\mathbf{w_1}},\vec{\mathbf{w_2}})\phi(\vec{\mathbf{w_1}})\chi(\vec{\mathbf{w_2}}). \end{equation} 
The variations in $\mathbf{\hat{V}}(\vec{\mathbf{w_1}},\vec{\mathbf{w_2}})$ lead to the development of entangled branches.

The amount of entanglement depends on the extent to which the variation in potential reshapes the wave functions. Typically, for microscopic systems the variation is relatively large. However, if one of the systems is macroscopic the large mass (and, hence, large spread in momentum) usually leads to an extremely narrow wave function. This means that the variation in $\mathbf{\hat{V}}(\vec{\mathbf{w_1}},\vec{\mathbf{w_2}})$ across it is extremely slight, and the effect on the wave function shape is minimal. Nevertheless, as long as it is regarded as a quantum system, there is some nonzero entanglement. It is worth emphasizing that this is true not only for a macroscopic system regarded as a measurement device, but also for a preparation apparatus. This has important consequences for the status of conservation laws in quantum theory.

Since there is always some residual entanglement with a preparation apparatus the notion of an elementary system being in an eigenstate of an observable is, at best, an idealized approximation. Because it is not in a factorizable state, one cannot attribute an eigenvalue to the measured (sub)system without reference to the larger system with which it is entangled. As pointed out in the Introduction, any assessment of measurement effects on conserved quantities must be based on the definition of those quantities that is used in the derivation of the conservation laws. As noted earlier the general schema for the derivation is given by Eq. \ref{1p1}, resulting in the definition for conserved quantities expressed in Eq. \ref{1p2}. 
This is the definition that will be used throughout this discussion.

 \section{Measurement, Collapse and Conservation Laws: Simple Illustration}
 \label{sec:3}

Before describing the stochastic collapse equation and demonstrating that it fully respects the relevant conservation laws throughout the collapse process it is worth illustrating some of the key ideas in qualitative terms with a simple example involving conservation of momentum.

When a photon is reflected from the surface of a mirror there is an exchange of momentum.\footnote{For the most part this argument will deal with the issue of conservation laws in \textit{nonrelativistic} quantum theory. However, for the essential aspects of entanglement exhibited in these simple sorts of exchange interactions, the relativistic nature of the photon and the technical complications in defining photon wave functions are irrelevant.} The states of both systems are changed from what they were prior to the interaction. This is very obvious for the photon, but the change in the state of the mirror is almost undetectable. If the mirror is a component of a beam-splitter that partially transmits the photon and partially reflects it, then the interaction results in some entanglement between the photon and the beam-splitter. Designate the branch of the photon that undergoes the reflection as $|\gamma_r\rangle$ and the transmitted branch as $|\gamma_t\rangle$. The beam-splitter state prior to the reflection can be labeled  $|b_0\rangle$, and the (very slightly) altered states brought about by the interaction can be labeled  $|b_r\rangle$ and $|b_t\rangle$. The resulting entangled state can be represented as 
$(1/\sqrt 2)(|b_r\rangle|\gamma_r\rangle + |b_t\rangle|\gamma_t\rangle )$ (assuming equal amplitudes for the two branches).

The photon and the beam-splitter states \textit{almost} factorize because there is very little difference between the resulting beam-splitter states:  $|\langle\;b_r\; | b_t\; \rangle | \; = \; 1-\delta$, where $\;\delta\, \ll \, 1.\; $ But the fact that the overlap between the beam-splitter states is not complete means that there is some nonzero entanglement. In order to illustrate this an explicit calculation of the entanglement between the photon and the beam-splitter is given in Appendix A (using the relative von Neumann entropy as an entanglement measure). For very small $\delta$ the entanglement can be approximated as $ \;(\delta/2)[1-\ln(\delta/2) ].$ Since entanglement is what mediates wave function collapse, if the reflected branch of the photon is subsequently detected, the detection collapses not only the photon wave function, but also the state of the mirror/beam-splitter which initially separated the two photon branches. If we include the detector, $|\mathcal{D}\rangle$, in the description, the collapse can be represented schematically as:  \newline 
$(1/\sqrt 2)(|b_r\rangle|\gamma_r\rangle + |b_t\rangle|\gamma_t\rangle ) 
\otimes|\mathcal{D}_0\rangle 
\;\; \Longrightarrow \;\;|b_r\rangle|\gamma'_r\rangle|\mathcal{D}_r\rangle$ (where $|\gamma'_r\rangle|\mathcal{D}_r\rangle$ represents the absorption of the reflected branch of the photon by the detector).

The difference in the states of the preparation apparatus is, of course, completely unobservable in practice. However, the point is that the difference in the momenta of the two photon branches originated in the exchange of momentum with the much larger system. These types of exchange create very small differences in the states of macroscopic systems. When the collapse of the wave function transfers all amplitude out of one state and into another, it eliminates any physical trace of the momentum correlation between the undetected branch of the photon and preparation apparatus, and it enhances the exchange that provided momentum to the branch that is detected. In other words, the momentum apparently lost from the branch that has disappeared is effectively transferred back to the system with which it had previously interacted, and the ``extra" momentum gained by the branch which has had its amplitude enhanced is paid for by the enhancement of the correlated state of the larger system(s) with which it has interacted. The total momentum of the combined system consisting of beam-splitter, photon, and detector is the same after the collapse of the wave function as it was before the interaction between  the photon and the beam-splitter.

The collapse effects will actually extend even further since macroscopic systems, such as $b$, will have also  interacted with their environments and become entangled with them. But this additional exchange of conserved quantities does not affect the essential point. There will also be residual entanglement relations between the measured system and systems with which it has interacted prior to encountering the preparation apparatus. In the example just described this would include the device that emitted the photon. The effects on the states of these ``pre-preparation" devices will naturally tend to decrease as one goes farther back in the entanglement chain. Later changes must be consistent with the pre-existing state, and, as these earlier systems develop new entanglement relations, the effects mediated by older, residual relations will diminish.\footnote{This point will be argued more formally in the Section 5.}

A couple of key aspects of this example should be noted. First, the manner in which the relevant quantities are conserved is \textit{nonlocal}, as is to be expected in situations involving wave function collapse. Initially, one might be concerned that such nonlocal changes in momentum or other conserved quantities might lead to superluminal signaling, since a measurement in one location can change the state of a distant system. The answer to this is that such changes can only be detected by a measurement on the system whose state has been changed. This ``second" measurement is itself capable of bringing about the collapse to the observed state; so no information about whether a distant measurement has occurred can be acquired.

Secondly, in order to determine whether a particular quantity is conserved in a given situation, one must track \textit{exchanges} of that quantity. This implies that both microscopic and macroscopic systems (including any preparation apparatus) must be described in strictly quantum terms, with no classical interface or boundary. It also implies that the only potentials in the Hamiltonian should be \textit{interaction} potentials. The stochastic collapse operator described in the next section is constructed from these interaction potentials.

\section{An Interaction-Induced Stochastic Collapse Equation}
\label{sec:4}

The projection postulate of von Neumann\cite{von_Neumann} states that measurements on quantum systems result in wave function collapse, but it gives very little indication of how this phenomenon is connected to fundamental physical processes. To deal with this explanatory gap a number of stochastic generalizations of the Schr\"{o}dinger equation have been proposed to explain collapse at the elementary level.

 Stochastic collapse equations begin with the (unitary) Schr\"{o}dinger equation in the form, 
 \begin{equation}\label{4p1}   
 |d\psi\,\rangle \,   = \, (-i/\hbar)\mathbf{\hat{H}}|\psi\,\rangle dt,
   \end{equation} 
  and then add nonunitary, stochastic modification terms. The additional terms are chosen to reproduce Schr\"{o}dinger evolution (to a very good approximation) for microscopic systems, but to yield reduction to an (approximate) eigenstate of the measured observable when macroscopic systems become involved. (``Measured" can refer to either actual measurements in a laboratory, or measurement by environmental interactions.) There is a substantial body of literature devoted to this subject; some of the main papers include\cite{Pearle_1976,Pearle_1979,Gisin_1984,GRW,Diosi_1,Diosi_2,Diosi_3,Gisin_c,GPR,Ghirardi_Bassi,Pearle_1,Brody_finite}.

 Stochastic collapse equations fall into several general categories depending on whether they are linear or nonlinear, and whether or not they preserve the norm of the state vector. This discussion will focus, primarily, on nonlinear, norm-preserving equations.

 A good general discussion of norm-preserving stochastic collapse equations is provided by Adler and Brun\cite{Adler_Brun}. One of the fairly general forms they arrive at can be represented as:
  \begin{equation}\label{4p2}   
 |d\psi\,\rangle \,   = \,  (-i/\hbar)\mathbf{\hat{H}}|\psi\,\rangle dt \, 
 - \, \frac{1}{2}\sum_k\mathbf{\hat{ \mathcal{B}}_k^\dagger}  \mathbf{\hat{ \mathcal{B}}_k}   
 |\psi\,\rangle dt \,  +\, \sum_k\mathbf{\hat{ \mathcal{B}}_k}|\psi\,\rangle d\xi_k 
 \end{equation} 
 As indicated above the first term on the right represents the ordinary Schr\"{o}dinger evolution, governed by the Hamiltonian, $\mathbf{\hat{H}}$. The third term involving $\mathbf{\hat{ \mathcal{B}}_k}$ represents the primary action of the stochastic operator(s), and the middle term, 
 $ \, - \frac{1}{2}\sum_k\mathbf{\hat{ \mathcal{B}}_k^\dagger}  
 \mathbf{\hat{ \mathcal{B}}_k}   |\psi\,\rangle dt, \,  $  
 can be understood as an adjustment to the length of the state vector that maintains the normalization. 
 The $\mathbf{\hat{ \mathcal{B}}_k}$ are nonlinear operators constructed from Lindblad operators,  $\mathbf{\hat{ L}_k}$, with 
 $ \,   \mathbf{\hat{ \mathcal{B}_k}} \;= \;  \mathbf{\hat{ L}_k} 
  - \langle \mathbf{\hat{ L}_k} \rangle. \, $  
This formulation uses the It$\hat{o}$ stochastic calculus, in which the $\xi_k $ are independent, Wiener (Brownian motion) processes. The mean of these processes is assumed to be zero, and the It$\hat{o}$ calculus rules determine the relationship among the differentials: 
 \begin{equation}\label{4p4}   
  d\xi^{*}_{j} d\xi_k = dt \delta_{jk}, \;\;\; dt d\xi_k = 0. 
  \end{equation} 
 Thus the $ d\xi_k$ have the units:  $ (dt)^{\frac{1}{2}}$.

 For this application the Lindblad operators are usually taken to be self-adjoint.  Typically, they correspond to a particular type of observable; the idea is that the wave function will eventually collapse to an eigenstate of that observable. What differentiates various approaches to stochastic collapse is the choice of operator type. The two approaches that have been most prominent seek to reduce $\psi$ either to an approximate position state, or to an energy eigenstate. The problem for position-based approaches in regard to conservation laws is that the narrowing of the wave function implies an increase in energy. It has been proposed that this energy could be supplied by a randomly fluctuating classical field\cite{Pearle_CL}, but this field is not directly observable. The energy eigenstate approach was originally suggested by Gisin\cite{Gisin_c} and has been developed extensively by Hughston and Brody\cite{Hughston_Brody}. In it the $\mathbf{\hat{ \mathcal{B}}_k}$ operators are constructed from the Hamiltonian. It does conserve energy on average, but, it has not been shown to do so in individual experiments.

In order to develop a collapse equation that respects conservation laws to the greatest extent possible we need to view wave function collapse as the selection of one entire branch in a system of entanglement relations, rather than as just the picking out of a particular eigenstate of the measured subsystem. We also need to focus on the interactions between subsystems that are responsible for the exchange of conserved quantities, and for the generation of the entanglement relations. These relations are what mediate collapse, but the way in which they do so depends on the definition of the collapse basis. The intimate link between the generation of entanglement and the exchange of conserved quantities through the interactions strongly suggests that the collapse basis should be defined in terms of these interactions.\footnote{Although its approach to the quantum measurement problem is fundamentally different from that advocated here, the decoherence program describes the collapse basis in essentially the same way.\cite{Zurek,Zurek_2}.} To implement this idea the collapse operators, $\mathbf{\hat{ \mathcal{B}}_k}$, will be constructed from the interaction potentials.

The effect of the operators should also be scaled to reflect the degree of entanglement that is generated by the interaction. As noted earlier, the entanglement between the measured system and the preparation apparatus is extremely small, while that between the measured system and measurement apparatus approaches a maximum prior to the complete collapse.\footnote{What distinguishes an apparatus used for preparation from one used for  measurement is in the way that measured subsystems interact with them. Since  instruments used to prepare a system are designed not to be affected to any noticeable extent by the interaction, a Schr\"{o}dinger equation describing the preparation process would characterize the interaction with a single potential and a very large apparatus mass. In contrast, measurement processes involve a chain or cascade of interactions between the measured system and very small components of the measurement instrument. } The scaling can be done by dividing the interaction potential by the combined mass of the interacting systems. We can begin by defining 

\begin{equation}\label{4p5}   
\hat{\mathcal{V}}_{jk}' \, = \, \mathbf{\hat{V}_{jk}}/(m_j+m_k), 
\end{equation} 
where $\mathbf{\hat{V}_{jk}}$ is the interaction potential between system $j$ and system $k$, and $m_j$ and $m_k$ are their masses. The term $ \, \hat{\mathcal{V}}_{jk}' \,$ needs to be further modified in order to maintain the dimensional consistency of the collapse equation. Since $ \mathbf{\hat{V}_{jk}} $ represents energy and the $ d\xi_k$ have the units,  $ (dt)^{\frac{1}{2}}$, the necessary consistency can be achieved by defining 
\begin{equation}\label{4p6}  
 \hat{\mathcal{V}}_{jk} \,=\, (\hat{\mathcal{V}}_{jk}'/c^2)\Gamma^{1/2}_{jk}, 
\end{equation} 
and setting  
\begin{equation}\label{4p7}  
 \mathbf{\hat{\mathcal{B}}_{jk}} \, = \, \hat{\mathcal{V}}_{jk}\, - \, <\hat{\mathcal{V}}_{jk}> .
\end{equation}
Here $c$ is the speed of light, and 
$\Gamma_{jk} \, \sim \, (1/dt)$ is the rate at which entanglement is generated by the interaction between system $j$ and system $k$.

The choice of $c$ to convert mass to energy is based on the fact that there simply is no other reasonable candidate. Since this discussion has been framed mainly in nonrelativistic terms, it does not carry its usual significance as a limiting speed. However, given the ratio of ordinary interaction energies to energies represented by expressions like $mc^2$, it does remind us that we expect the collapse-inducing effect of individual elementary interactions to be very small. The definition of the parameter, $\Gamma_{jk}$, is motivated by idea that collapse is induced by \textit{entangling} interactions. As described in Section 2 entanglement is a generic result of the interaction between two systems, and this is most obvious for previously unentangled systems. However, after two systems have become bound into a stationary state, no additional entanglement is generated. By defining $\Gamma_{jk}$ in this way, the time parameter(s) get determined in a completely nonarbitrary manner, and stationary states remain undisturbed by the stochastic process. Finally, since many interactions will be taking place at any given time, and to insure the overall consistency of the collapse process we can combine the individual collapse operators to form a single, universal operator: 
 \begin{equation}\label{4p8}
  \mathbf{\hat{\mathcal{B}}} \, = \, 
 \sum_{j<k} \,\mathbf{\hat{\mathcal{B}}_{jk}}, 
 \end{equation}   
 with a single Wiener process, $\xi$.
 In the collapse equation proposed here the type of state to which the wave function collapses is determined by the particular arrangement of interactions designed into the measurement instrument. The observable to which the relevant states correspond can vary, but it is worth noting that since the interaction potentials are distance-dependent, measurement results typically approximate position eigenstates.

 The argument for this version of the collapse equation has been based on the possibility of maintaining the conservation laws to the greatest extent possible. Essentially, the same formulation is argued for separately in \cite{Gillis_2} based on the no-superluminal-signaling principle. That work provides a more comprehensive discussion of the issues surrounding quantum measurement, and, in particular, it addresses the problem of reconciling the nonlocal nature of wave function collapse with relativity. It also describes how the equation insures collapse to one of the expected outcomes with the correct probabilities.\footnote{A more general discussion on this point is contained in \cite{Adler_Brun}.}

\section{Assumptions and Initial Conditions}
\label{sec:5}

Conserved quantities are exchanged through interactions, and those interactions generate entanglement. In nonrelativistic theory interactions are represented by potentials with nonzero gradients. To insure strict conservation these must be functions of relative distance only:  
\begin{equation}\label{5p1}  
 \mathbf{\hat{V}}(\mathbf{w_1},\mathbf{w_2}) \;  = \;   \mathbf{\hat{V}}(  |  \mathbf{w_1} \, - \, \mathbf{w_2} | ) . 
\end{equation} 
This has the immediate consequence that: 
\begin{equation}\label{5p2} 	
  \mathbf{\nabla_1} \mathbf{\hat{V}}(\mathbf{w_1},\mathbf{w_2})  \; 
= \; - \mathbf{\nabla_2} \mathbf{\hat{V}}(\mathbf{w_1},\mathbf{w_2}). 
\end{equation}
This relation is crucial for maintaining conservation laws for both ordinary Schr\"{o}dinger evolution and for the stochastic generalization to be discussed here.

Given the nonzero gradient of the potential, the generation of entanglement (between previously unentangled systems) by the interactions is evident from the potential energy term in the Schr\"{o}dinger equation as described in Section 2. Since the interaction potentials vary continuously with position, they cause the wave function to deviate from a simple product structure across any local region in configuration space in which they are present, no matter how small that region is. This implies that wave function collapse never completely eliminates entanglement, since even after collapse, there is always some nonzero spread to the wave function.

Along with the stochastic collapse equation described in the previous section, the proof offered here relies on one additional assumption about initial conditions. This assumption requires some explanation and justification. As pointed out at the end of Section 3, due to the fact that there is always some entanglement all relevant systems must be described in quantum terms. There can be no classical boundary. This raises a problem concerning the specification of initial conditions. Ordinarily, the stipulation of these conditions assumes a classical boundary (at least, implicitly), and eliminates any consideration of interactions across this boundary. The effect of this implicit assumption is to ignore the small amount of residual entanglement between the microscopic,  ``quantum" system under consideration and its macroscopic, ``classical" surroundings. But, it is this small amount of entanglement that is essential to understanding what happens to conserved quantities. So, to properly deal with the issue of conservation, one must ask what is the least arbitrary assumption about the initial state of the system. In particular, it is important to consider how conserved quantities are ``initially" distributed among the entangled branches of the total system.

In order to analyze the effects of all interactions, including those between microscopic and macroscopic systems, I will consider a total system consisting of a large number of elementary systems that have been interacting for an extended time. In this sort of situation it is reasonable to assume that the properties of any elementary system are determined to an arbitrarily high degree by the interactions that it has undergone in the past. This means that the particular state of such a (sub)system in the distant past is essentially irrelevant to its current state. The entanglement structure of that state and the way in which conserved quantities are distributed among the entangled branches are (almost) entirely due to exchanges of those quantities through interactions with other elementary systems.

Entangled states can, of course, be decomposed into a variety of different bases. The decomposition to be studied here is the one determined by the interaction potentials.\footnote{This ``interaction" basis can, of course, change as the system evolves. However, in measurement situations the organized way in which the interactions are arranged helps to insure that a particular basis is picked out. In more general situations this basis essentially coincides with the decoherence basis (discussed by Zurek\cite{Zurek})  which leads to separation into well defined branches.} As noted in the previous section these potentials are responsible for both the generation of entanglement and the exchange of conserved quantities. This makes the decomposition into the entangled branches defined by interactions the natural choice for the collapse basis. The critical assumption that will be made here is that conserved quantities are shared equally (up to normalization) by the entangled branches. The distribution of a quantity such as momentum among the entangled subsystems that are included in the branch will vary. But, the total quantity, calculated across each entangled branch, will be proportional to the squared amplitude of that branch.

To spell out this assumption in mathematical terms let 
$ \psi({\mathbf{w_1,w_2,...,w_j,w_k,...}} ) $ represent the wave function of the total system in configuration space. It was pointed out above that the nonzero gradients of the position-dependent interaction potentials generate entanglement  in the wave function across even the smallest of regions in configuration space. This means that, at the most fine-grained level, for any entangled subsystem, $j$, associated with the variable, $\mathbf{w_j}$, each distinct value of $\mathbf{w_j}$ defines a distinct entangled branch of the overall wave function in the basis defined by the interactions occurring at that stage of evolution. (This interaction basis will, of course, evolve over time, and the labeling of entangled branches by a particular value of a coordinate will change as the wave function evolves.) This labeling by distinct coordinate values allows one to identify entangled branches simply by setting one of the configuration space variables to a fixed value:\footnote{Subsystems that are not entangled with subsystem $j$ (those that are in a product relationship with it) will, by default, also have nonzero amplitudes at $\mathbf{w_j} \; = \; \mathbf{a_j}.$ But, this does not affect the analysis here.}.  
\begin{equation}\label{5p3}   
  \psi_{\mathbf{a_j}}  \; = \; \psi({\mathbf{w_1,w_2,...,w_j = a_j,w_k,...}} ).
\end{equation} Suppose that we fix 
$ \mathbf{w_1}$. The normalized value of a conserved quantity such as x-momentum, 
$ (-i\hbar(\partial{\psi}/\partial{x}))  $, within an entangled branch can then be represented as:    
\begin{equation}\label{5p4}   
   \dfrac{\int \, d\mathbf{w_2}d\mathbf{w_3}...\; \psi^*({\mathbf{a,w_2,w_3,...}})
	[\sum_j (-i\hbar(\partial{\psi({\mathbf{a,w_2,w_3,...}})}/\partial{x_j}))]}
{ \int \, d\mathbf{w_2}d\mathbf{w_3}...\; \; \psi^*({\mathbf{a,w_2,w_3,...}})\psi({\mathbf{a,w_2,w_3,...}})}.
\end{equation}

If the integrations in the numerator and denominator were not restricted by the condition, $ \mathbf{w_1} \, = \, \mathbf{a} $, the expressions could be more compactly represented with the Dirac bra-ket notation as: $ \langle\psi|\hat{\mathbf{Q}}|\psi\rangle$ and 
$ \langle\psi|\psi\rangle$, where $\hat{\mathbf{Q}}$ represents the operator corresponding to a conserved quantity (x-momentum in this particular case). In order to simplify the notation I will adapt the Dirac convention to abbreviate the expressions as follows. The integrations \textit{with} the restriction, $ \mathbf{w_j} \, = \, \mathbf{a}$, will be represented as: $ \langle\psi|\hat{\mathbf{Q}}|\psi\rangle_{ \mathbf{a_j} }$, and 
$ \langle\psi|\psi\rangle_{ \mathbf{a_j}  }$. This allows us to define:  
\begin{equation}\label{5p5}   
 \mathbf{q}_{ \mathbf{a_j} } \; \equiv \;(\, \langle\psi|\hat{\mathbf{Q}}|\psi\rangle_{ \mathbf{a_j} }\,) \; / \; ( \, \langle\psi|\psi\rangle_{ \mathbf{a_j} } \, ). 
\end{equation}
(Similarly, a restriction of two variables, $\mathbf{w_j}$ and $\mathbf{w_k}$, to fixed values can be represented as  $ \langle\psi|\hat{\mathbf{Q}}|\psi\rangle_{ \mathbf{a_j,b_k} }$, and $ \langle\psi|\psi\rangle_{ \mathbf{a_j,b_k}  }$.)

The assumption about conserved quantities being equally shared among entangled branches  can then be stated as follows. For any two distinct fixed values, 
$ \mathbf{w_j} \, = \, \mathbf{a} $,   and $ \mathbf{w_j} \, = \, \mathbf{a'} $, 
$ \mathbf{q}_{  \mathbf{a_j} } \; 
= \; \mathbf{q}_{  \mathbf{a_j'} } $.

This assumption is based on the stipulation that the systems have been interacting for an extended time. This implies that any elementary system will have interacted with a very large number of other such systems. Therefore, it will have exchanged conserved quantities and become entangled with a large number of these other systems. In any particular entangled branch the values of conserved quantities associated with a specific elementary system will vary from the average for that system. However, since the entangled branches will include a very large number of elementary systems, basic probability considerations indicate that the deviations from the average over the entire branch will tend to cancel out. Therefore, 
total amounts of momentum, angular momentum, and energy in each branch should  approach average values.

 Up to this point I have argued as follows. Wave function collapse is induced by entangling interactions. It is mediated by entanglement relations. The collapse basis is defined by the interactions that generate the entanglement. Some entanglement relations always survive collapse.  Therefore, collapse should be viewed as picking out an entire entangled branch rather than an eigenstate of a small subsystem. Given a system with a large enough number of elementary subsystems that have undergone a sufficient number of interactions, the value of a conserved quantity calculated across the entire branch approaches the average value  for the total system (up to normalization of the branch). The conclusion that follows from this general outline is that the value of this quantity does not change when one of the branches is selected by the collapse. What remains to be shown is that the stochastic collapse equation described in Section 4 guarantees conservation in this manner.	\cite{Bartsch_Gemmer}

 \section{Conservation of Momentum and Angular Momentum}
 \label{sec:6}
 
 Stochastic collapse equations operate by transferring amplitude among the various branches of the wave function until all of the amplitude resides in one branch. The strategy for proving that conservation is guaranteed by the equation described in Section 4 is to show that the change in the relevant quantities in each branch induced by the collapse process is directly proportional to the change in the squared amplitude of the branch. In other words, the idea is to show that the change in the normalized value of the quantity defined above in Eq. \ref{5p5} is zero.

  A general form for norm-preserving stochastic collapse equations was presented in Section 4. With a single Lindblad operator it can be represented as: \newline 
 \begin{equation}\label{6p1} 
   |d\psi\,\rangle \, = \, -\frac{i}{\hbar}\mathbf{\hat{H}}|\psi\,\rangle   dt \, - \, \frac{1}{2}\mathbf{\hat{ \mathcal{B}}^\dagger}
  \mathbf{\hat{ \mathcal{B}}}  |\psi\,\rangle  dt \,  +\, \mathbf{\hat{ \mathcal{B}}}|\psi\,\rangle  d\xi. 
  \end{equation}   
   The bra-ket notation is useful when discussing the vector characteristics of the wave function such as linearity and normalization. However, in what follows it is necessary to examine variations in the wave function in local regions of configuration space. Since these variations cancel when an integration is done over the entire wave function, we need to use the more general form of the equation without the vector notation: 
  \begin{equation}\label{6p2} 
     d\psi \, = \, -(\frac{i}{\hbar}\mathbf{\hat{H}}) \psi\, dt \, 
   - \, \frac{1}{2}\mathbf{\hat{ \mathcal{B}}^\dagger}
   \mathbf{\hat{ \mathcal{B}}}  \psi\,  dt \,  +\, \mathbf{\hat{ \mathcal{B}}} \psi\,  d\xi.   
   \end{equation} 
 The change induced in $ \psi\ $ by the stochastic process implies a corresponding change in the adjoint wave function: 
 \begin{equation}\label{6p3} 
    d\psi^* \; = \; (+\frac{i}{\hbar}\mathbf{\hat{H}} ) \psi^* \,  dt \, - \, 
   (\frac{1}{2}\mathbf{\hat{ \mathcal{B}}}
  \mathbf{\hat{ \mathcal{B}}^\dagger}) \psi^*  \,  dt \,
  + \, \mathbf{\hat{ \mathcal{B}}^\dagger} \psi^*  \,   d\xi^*.  
  \end{equation} 
  The change in the product,  $\psi^*\psi$, can be represented schematically as: \begin{equation}\label{6p4} 
 d(\psi^*\psi) \; = \; (\psi^* + d\psi^*)(\psi + d\psi) \, - \, (\psi^*\psi) \; = \;   d\psi^* \psi \, + \psi^*  d\psi \,  +  d\psi^* d\psi. 
\end{equation}

  Ordinarily, differentials are computed by discarding terms that are higher than first order in small increments of the relevant variables. However, since the Wiener process, $ \xi(t) $, has the general character of a random walk, it varies as $ \sqrt{t}$. Given the It$\hat{o}$ calculus rule, Eq. \ref{4p4} ($d\xi^{*} d\xi = dt $),  we must retain terms with quadratic variations in  $d\xi$. The changes that this forces in the rules for differentiation in the It$\hat{o}$ calculus affect both the chain rule and the (Leibniz) product rule.\footnote{The term, $ dt d\xi$, is of more than first order in $dt$ which is why the It$\hat{o}$ calculus rules state that $dt d\xi = 0$.}

 Collecting terms and eliminating those that are higher than first order in $dt$ results in the following expression for the product,  $\psi^*\psi$:  
 \begin{equation}\label{6p5}  
 \begin{array}{ll} 
 d(\psi^*\psi) \; = \; 
 - \psi^* (\frac{1}{2}\mathbf{\hat{ \mathcal{B}}}
 \mathbf{\hat{ \mathcal{B}}^\dagger})  \, \psi  \, dt \,  
 - \psi^* (\frac{1}{2}\mathbf{\hat{ \mathcal{B}}^\dagger}
 \mathbf{\hat{ \mathcal{B}}})  \, \psi  \, dt \, 
  + \, \psi^* \, \mathbf{\hat{ \mathcal{B}}^\dagger}  \mathbf{\hat{ \mathcal{B}}} 
 \psi  \, d\xi^*d\xi \,   
   & \\    
 \; \;\; \;   \; \;\; \; \; \;  \; \;\; \;  \;\; \;\; + \; \; 
  \psi^* \,    \mathbf{\hat{ \mathcal{B}}^\dagger}  \psi    \, d\xi^* \, +\,  \psi^* \,    \mathbf{\hat{ \mathcal{B}}} \psi  \, d\xi. 
   \end{array}
  \end{equation}
 Since $\mathbf{\hat{ \mathcal{B}}} $ has been constructed from a self-adjoint Lindblad operator, $\mathbf{\hat{ \mathcal{B}}^\dagger} \, = \, \mathbf{\hat{ \mathcal{B}}}$. This condition can be used along with Eq. \ref{4p4} to show that the first three terms on the right cancel. Thus, we arrive at:
 \begin{equation}\label{6p6}     
 \;\;\;   d (\psi^* \psi) \; = \; 
 \psi^* \,    \mathbf{\hat{ \mathcal{B}}^\dagger}  \psi  \, d\xi^* 
 \, + \, \psi^*  \,    \mathbf{\hat{ \mathcal{B}}} \psi \, d\xi.  
 \end{equation}  
 (The adjoint notation for the term, $ \, \psi^* \,    \mathbf{\hat{ \mathcal{B}}^\dagger}  \psi  \, d\xi^*, \,$ has been temporarily retained as a reminder of the order in which other operators are to be inserted into the expression.)

In order to show that the normalized value of a quantity in a particular branch, $\mathbf{q}_{ \mathbf{a_j} },$ as defined in Eq. \ref{5p5} 
is unchanged by the stochastic process we need to show that:  
\begin{equation}\label{6p7}   
     \langle \psi + d\psi \, | \, \hat{\mathbf{Q}} \,| \, \psi + d\psi \, \rangle_{ \mathbf{a_j} }\, \; / \;  \, \langle \psi + d\psi | \psi + d\psi\rangle_{ \mathbf{a_j} }  \; = \;  \langle\psi|\hat{\mathbf{Q}}|\psi\rangle_{ \mathbf{a_j} }\, \; / \;  \, \langle\psi|\psi\rangle_{ \mathbf{a_j} }. 
 \end{equation}  
  It is clear that this will hold if we can show that, at each point in configuration space, 
\begin{equation}\label{6p8}   
    d(\psi^*\hat{\mathbf{Q}} \psi) \; / \;   d (\psi^* \psi) \; = \;  \psi^*\hat{\mathbf{Q}} \psi \; / \;  \psi^* \psi. 
\end{equation}

  Following the pattern that was used to compute $d(\psi^* \psi) $ above,  the effect of the stochastic process on 
  $ (\psi^* \, \mathbf{\hat{Q}} \, \psi)$ can be calculated as 
\begin{equation}\label{6p9}   
 \begin{array}{ll} 
   d (\psi^* \, \mathbf{\hat{Q}} \,\psi) \; = \;  -\psi^* (\frac{1}{2}  \mathbf{\hat{  \mathcal{B}}}
  \mathbf{\hat{  \mathcal{B}}^\dagger}) \, \,\mathbf{\hat{Q}} \, \psi \, dt \; - \;
  \psi^* \, \hat{\mathbf{Q}} \, (\frac{1}{2}\mathbf{\hat{  \mathcal{B}}^\dagger}
  \mathbf{\hat{  \mathcal{B}}}) \, \psi \, dt \,   
 
   + \; \psi^* \, \mathbf{\hat{  \mathcal{B}}^\dagger}  \hat{\mathbf{Q}}  \mathbf{\hat{  \mathcal{B}}} \,  \psi   \, d\xi^*d\xi
 \; 
  & \\    
  \; \;\; \;   \; \;\; \; \; \;  \; \;\; \;   \; \;\; \;   \;\; \;\; + \; \; 
   \psi^* \,  \mathbf{\hat{  \mathcal{B}}^\dagger}  \hat{\mathbf{Q}} \, \psi  \, d\xi^* \; + \;\psi^* \,  \hat{\mathbf{Q}}   
  \mathbf{\hat{  \mathcal{B}}} \,  \psi  \, d\xi.  
  \end{array}
   \end{equation} 
Dropping the adjoint notation and again invoking  $d\xi^*d\xi \, = \, dt$, the expression on the right becomes:  
\begin{equation}\label{6pa} 
   \;     \psi^*( -\frac{1}{2} [  \mathbf{\hat{  \mathcal{B}}}
\mathbf{\hat{  \mathcal{B}}}  \mathbf{\hat{Q}}  \, + \,
\hat{\mathbf{Q}}  \mathbf{\hat{  \mathcal{B}}}
\mathbf{\hat{  \mathcal{B}}} ]    \;   
+ \; \mathbf{\hat{  \mathcal{B}}}  \hat{\mathbf{Q}}  \mathbf{\hat{  \mathcal{B}}} \, )\psi   \, dt \; \;   + \;  \psi^* \,  \mathbf{\hat{  \mathcal{B}}}  \hat{\mathbf{Q}} \, \psi  \, d\xi^* \; + \;\psi^* \,  \hat{\mathbf{Q}}   
\mathbf{\hat{  \mathcal{B}}} \, \psi  \, d\xi. 
\end{equation} 
Above,  we saw that
\begin{equation}\label{6pb}  d (\psi^* \psi) \; = \; 
\psi^* \,    \mathbf{\hat{ \mathcal{B}}}  \psi  \, d\xi^* 
\, + \, \psi^*  \,    \mathbf{\hat{ \mathcal{B}}} \psi \, d\xi  \; = \; 
(\psi^* \, \mathbf{\hat{ \mathcal{B}}} \,  \psi)  \,     (d\xi^* \, + \, d\xi). 
 \end{equation}    
So, what needs to be demonstrated is that: 
\begin{equation}\label{6pc}
  d (\psi^* \hat{\mathbf{Q}} \psi) \; = \;  
(\psi^* \, \mathbf{\hat{ \mathcal{B}}} \, \hat{\mathbf{Q}}\,  \psi)  \,     (d\xi^* \, + \, d\xi). 
\end{equation} 
This can be done by showing that the relationship, 
$ \,   \hat{\mathbf{Q}}   \mathbf{\hat{  \mathcal{B}}} \, \psi \; = \;   
 \mathbf{\hat{  \mathcal{B}}} \, \hat{\mathbf{Q}} \psi, \, $
 holds as an identity, at each point in configuration space - not simply as an operator relationship. 
 This will imply that the first three terms above (multiplying $dt$) cancel, and that 
\begin{equation}\label{6pe} 
  \;\psi^* \, \hat{\mathbf{Q}} \mathbf{\hat{ \mathcal{B}}} \, \psi \, d\xi  \; 
  = \;  \;\psi^* \,\mathbf{\hat{ \mathcal{B}}}  \hat{\mathbf{Q}} \, \psi \, d\xi, 
\end{equation} 
 yielding the desired proportionality.

Let us begin with the momentum operator,
\begin{equation}\label{6pf} 
 \mathbf{\hat{P}} \, \equiv \, 
 -i \hbar \sum_n \,\mathbf{\hat{\nabla}_n}. 
 \end{equation} 
Recall the definition of $ \,  \mathbf{\hat{\mathcal{B}}} \, \equiv \, 
\sum_{j<k} \,\mathbf{\hat{\mathcal{B}}_{jk}}\, $ from Eq. \ref{4p8}, and the associated definitions of $  \mathbf{\hat{\mathcal{B}}_{jk}} \, = \, \hat{\mathcal{V}}_{jk}\, - \, <\hat{\mathcal{V}}_{jk}> $ Eq. (\ref{4p7}),  $ \hat{\mathcal{V}}_{jk} \, = \, (\mathbf{\hat{V}_{jk}}/[(m_j+m_k)c^2])\Gamma^{1/2}_{jk} $ Eq. (\ref{4p6}), and  $\Gamma_{jk} $  (the rate at which entanglement is generated by the interaction between system $j$ and system $k$). Since, at any particular time, the only factor in $ \mathbf{\hat{\mathcal{B}}_{jk}} $ with any coordinate dependence is 
  $\mathbf{\hat{V}_{jk}}$, we can expand 
 $ \mathbf{\hat{P}} \, \mathbf{\hat{\mathcal{B}}_{jk}} \, \psi,  \;$ as: 
\begin{equation}\label{6pg} 
 \{ (-i\hbar/[(m_j+m_k)c^2]) \Gamma^{1/2}_{jk}   \sum_n \,\mathbf{\hat{\nabla}_n} \, \mathbf{\hat{V}_{jk}} \} \, * \, \psi \; + \; \mathbf{\hat{\mathcal{B}}_{jk}}*(\mathbf{\hat{P}}  \, \psi).   
 \end{equation} 
To establish the proportionality that we want, it is necessary to show that the first term is zero. Since 
$\mathbf{\hat{\nabla}_l} \, \mathbf{\hat{V}_{jk}}  
\; = \; 0 $ for $ l \, \neq \, j,k, \,$ this reduces to showing that 
$(\mathbf{\hat{\nabla}_j} + \mathbf{\hat{\nabla}_j}) \, \mathbf{\hat{V}_{jk}}  
\; = \; 0. $ Conservative, two-system interaction potentials must obey the condition Eq. \ref{5p1} ($ \mathbf{\hat{V}_{jk}} \;  = \;   \mathbf{\hat{V}}(  |  \mathbf{w_1} \, - \, \mathbf{w_2} | ), $), which implies Eq. \ref{5p2} ($  \mathbf{\nabla_j} \mathbf{\hat{V}_{jk}}  \; = \; - \mathbf{\nabla_k} \mathbf{\hat{V}_{jk}} $). Hence, the first term in Eq. \ref{6pg} vanishes, and so we see that momentum is strictly conserved.

The derivation of conservation of orbital angular momentum is slightly more involved, but it follows a generally similar pattern. The calculation will be illustrated for the x-component of angular momentum, $\mathbf{\hat{L}_x}$.
After expanding 
$\mathbf{\hat{L}_x}  \mathbf{\hat{  \mathcal{B}}}\, \psi $ as we did above for 
$\mathbf{\hat{P}}  \mathbf{\hat{ \mathcal{B}}}\, \psi, $ what we have to show is that: 
\begin{equation}\label{6pi} 
 -i\hbar[((y_j\partial{\mathbf{\hat{V}_{jk}}}/\partial{z_j}  -  z_j\partial{\mathbf{\hat{V}_{jk}}}/\partial{y_j}) 
\, + \, (y_k\partial{\mathbf{\hat{V}_{jk}}}/\partial{z_k}  -  z_k\partial{\mathbf{\hat{V}_{jk}}}/\partial{y_k})]  \; = \; 0.
\end{equation}

We first define 
\begin{equation}\label{6pj} 
 r \; \equiv \; + \sqrt{(x_j-x_k)^2 +   (y_j-y_k)^2 + (z_j-z_k)^2 } 
\end{equation} 
so that 
\begin{equation}\label{6pj} 
 \mathbf{\hat{V}_{jk}}  \, = \, \mathbf{\hat{V}_{jk}}(r) ; \;\;\;\;\;
 \partial{\mathbf{\hat{V}_{jk}}}/\partial{z_j}   \, = \, (\partial{\mathbf{\hat{V}_{jk}}}/\partial{r})  \partial{r}/\partial{z_j}, 
\end{equation}
etc. Taking the partial derivatives of $r$ with respect to the coordinates, and inserting them into the expression yields:  
\begin{equation}\label{6pk}
 (-i\hbar)(\partial{\mathbf{\hat{V}_{jk}}}/\partial{r} )
[y_j(z_j-z_k)/r -  z_j(y_j-y_k)/r 
\, + \, y_k(z_k-z_j)/r  -  z_k(y_k-y_j)/r] . 
\end{equation}
Rearrangement of the terms produces the zero result: 
\begin{equation}\label{6pl}
 (-i\hbar)(\partial{\mathbf{\hat{V}_{jk}}}/\partial{r})(1/r)
[ \, (y_j-y_k)(z_j-z_k) \,  - \,  (z_j-z_k)(y_j-y_k) \, ]  \; = \; 0.
\end{equation}
So orbital angular momentum is also strictly conserved.

In order to get an overall picture of what has been shown one should note that, although the terms involving derivatives of $\mathbf{\hat{  \mathcal{B}}_{jk}} $ operate only on the $j$ and $k$ subsystems, the purely multiplicative terms involving $\mathbf{\hat{  \mathcal{B}}_{jk}} $ act to rescale the amplitudes of \textit{all} systems across the hyperplane defined by $\mathbf{w_j} = \mathbf{a_j}$ and $\mathbf{w_k} = \mathbf{b_k}$. The overall effect of the nonlocal changes induced by the operator,  
$ \mathbf{\hat{\mathcal{B}}} \, \equiv \, 
\sum_{j<k} \,\mathbf{\hat{\mathcal{B}}_{jk}},$  is simply to rescale the quantities of momentum and orbital angular momentum in exact proportion to the rescaling of the squared amplitude, insuring the strict conservation of these quantities throughout the collapse process.

One reason that the proofs of strict conservation for momentum and orbital angular momentum go through fairly easily is that they are defined in terms of first-order differential operators, since they are vector (or axial vector) quantities. Since it is represented as a vector quantity in nonrelativistic theory momentum translates fairly readily into the spatial component of a four-vector in relativity, and so its relativistic definition is unchanged. A similar point can be made about orbital angular momentum which is just the product of two vectors.

The situation with energy is much different. In nonrelativistic theory it is a scalar quantity, defined in terms of a second-order differential operator. In  relativistic theory it becomes the timelike component of a four-vector and the associated operator has a completely different character. This situation limits the accuracy of the energy calculations in nonrelativistic theory. Although the relativistic corrections are quite small at low speeds, they do become apparent at a sufficiently high level of precision. These factors complicate the discussion of conservation, and for this reason it is treated separately, in the next section.

\section{Energy Conservation and Relativistic Considerations}
\label{sec:7}

The main issue concerning energy conservation for stochastic collapse equations is that the localization of a particle in a small region narrows its wave function, and increases it derivatives, suggesting an apparent increase in energy. In the previous section we saw that, despite the increase in the first derivatives, momentum and angular momentum are conserved due to compensating changes in systems with which the particle has previously interacted. The most critical task in this section is to demonstrate that similar correlated changes across each entangled branch also eliminate the principal apparent deviation from energy conservation.

It is also important that the collapse equation described in Section 4 strictly conserve energy in all situations in which standard nonrelativistic quantum theory \textit{correctly} predicts energy conservation. To do this we must first note how limited these situations are. Since energy is the quantity associated with the Hamiltonian evolution operator the standard theory predicts that it is conserved in all situations in which the Hamiltonian is independent of time. The problem with this prediction is that the quantity that is conserved is defined as the sum of nonrelativistic potential and kinetic energy. This very restrictive definition excludes changes in kinetic energy associated with relativistic changes in mass, as well as conversion of energy to other forms such as radiation. Strictly speaking, the only situations in which the quantity described by the nonrelativistic formula is conserved are those in which there is zero change in kinetic energy. The only cases that fit this criterion are stationary states and
those in which there is no interaction.

In this limited set of cases the stochastic collapse equation of Section 4 reduces to the nonrelativistic Schr\"{o}dinger equation. The stochastic operator, 
$ \mathbf{\hat{\mathcal{B}}},$ obviously reduces to zero when there are no interactions. It also vanishes for stationary states since the entanglement rate, $\Gamma_{jk},$ goes to zero. One can see this by noting that steady state density matrices, $\rho$, do not change with time. Because they are constant, all entanglement measures based on them are also constant in time.

Whenever there are interactions (other than those that bind subsystems in stationary states) there are changes in kinetic energy. In all of these situations there are small deviations from strict conservation of the nonrelativistic forms of energy. In precisely these cases the stochastic collapse equation also generates small deviations for these forms of energy. We will see below that the deviations in both cases are of approximately the same magnitude. I argue that in neither case should the deviations be taken as evidence of genuine violations of energy conservation. Rather, they should be understood as indicating the limits of a nonrelativistic formulation. First, however, we will show how interaction-induced collapse eliminates the major problem with energy conservation.

 In the previous section it was shown that, over any region of configuration space, the change induced in the squared amplitude of the wave function by the stochastic process is given by  Eq. \ref{6pb}  
  ($  d (\psi^* \psi) \; = \;  \psi^* \,    \mathbf{\hat{ \mathcal{B}}}  \psi  \, d\xi^* 
\, + \, \psi^*  \,    \mathbf{\hat{ \mathcal{B}}} \psi \, d\xi \; = \; 
(\psi^* \,   \mathbf{\hat{ \mathcal{B}}} \,   \psi)  \,    (d\xi^* \, + \, d\xi). \; $
The associated effect on a quantity, $\mathbf{q}$, is expressed by \newline 
Eq. \ref{6pa} 
 ($   \psi^*( -\frac{1}{2} [  \mathbf{\hat{  \mathcal{B}}}
\mathbf{\hat{  \mathcal{B}}}  \mathbf{\hat{Q}}  \, + \,
\hat{\mathbf{Q}}  \mathbf{\hat{  \mathcal{B}}}
\mathbf{\hat{  \mathcal{B}}} ]    \;   
+ \; \mathbf{\hat{  \mathcal{B}}}  \hat{\mathbf{Q}}  \mathbf{\hat{  \mathcal{B}}} \, )\psi   \, dt \; \;   + \;  \psi^* \,  \mathbf{\hat{  \mathcal{B}}}  \hat{\mathbf{Q}} \, \psi  \, d\xi^* \; + \;\psi^* \,  \hat{\mathbf{Q}}   
\mathbf{\hat{  \mathcal{B}}} \, \psi  \, d\xi. \;$ )
\newline
For the cases of momentum and angular momentum it was possible to show that this expression simplified to Eq. \ref{6pc} ( 
$ \, d (\psi^* \, \mathbf{\hat{Q}} \, \psi)  \; = \; 
(\psi^* \, \mathbf{\hat{ \mathcal{B}}} \,  \mathbf{\hat{Q}} \, \psi) \,(d\xi^* \, + \, d\xi)$), 
demonstrating strict conservation of the quantities in question. For the reasons discussed above this exact proportionality does not hold in the case of energy. What needs to be determined is how close we can come to it when 
 \begin{equation}\label{7p1} 
  \,  \hat{\mathbf{Q}}  \; = \; \mathbf{\hat{H}}  
 \; = \;  -\frac{\hbar^2}{2m} \mathbf{\nabla}^2\, + \, \mathbf{\hat{V}} . 
 \end{equation}
Let us first examine the term, 
 \begin{equation}\label{7p2} 
 \psi^* \,   \hat{\mathbf{H}}  \mathbf{\hat{  \mathcal{B}}}  \, \psi  \, d\xi.
 \end{equation}

Since the stochastic operator, $ \mathbf{\hat{  \mathcal{B}}}, $
is constructed from the potential energy operator it is obvious that the following relationship holds at every point in configuration space: 
 \begin{equation}\label{7p3}
\mathbf{\hat{V}}  \mathbf{\hat{  \mathcal{B}}} \psi \; = \; \mathbf{\hat{  \mathcal{B}}}  \mathbf{\hat{V}} \psi.   
\end{equation}
As before, $  \mathbf{\hat{  \mathcal{B}}}  $ can be decomposed into individual pairwise factors, $ \mathbf{\hat{\mathcal{B}}_{jk} } $. For each  
$ \mathbf{\hat{\mathcal{B}}_{jk} } $ the Hamiltonian, $\mathbf{\hat{H}}, $ can be written as: 
 \begin{equation}\label{7p4}
  \mathbf{\hat{H}'} \, + \, [-\frac{\hbar^2}{2m} (\mathbf{\nabla_j}^2 \, + \, \mathbf{\nabla_k}^2)]. 
\end{equation} 
 Clearly, 
 \begin{equation}\label{7p5}
\mathbf{\hat{H}'}  \mathbf{\hat{\mathcal{B}}_{jk}} \psi \; = \; \mathbf{\hat{\mathcal{B}}_{jk}}  \mathbf{\hat{H}'} \psi.
\end{equation}  
So $\mathbf{\hat{H}}  \mathbf{\hat{\mathcal{B}}_{jk}} \psi $ becomes 
  $ \, \mathbf{\hat{\mathcal{B}}_{jk}}  \mathbf{\hat{H}'} \psi \;  -\frac{\hbar^2}{2m} (\mathbf{\nabla_j}^2 \, 
  + \, \mathbf{\nabla_k}^2) \, \mathbf{\hat{  \mathcal{B}}_{jk}} \psi. $ 
   
After expanding the second part of this expression, and rearranging we get: 
 \begin{equation}\label{7p6}
  \begin{array}{ll}  
 \{\mathbf{\hat{\mathcal{B}}_{jk}}  \mathbf{\hat{H}'} \psi \; 
  + \;     \mathbf{\hat{  \mathcal{B}}_{jk}} 
  [-\frac{\hbar^2}{2m} (\mathbf{\nabla_j}^2 \psi \, + \, \mathbf{\nabla_k}^2 \psi)]  \}  
   & \\    
    + \;   [-\frac{\hbar^2}{2m} (\mathbf{\nabla_j}^2  \mathbf{\hat{  \mathcal{B}}_{jk}} \, + \, \mathbf{\nabla_k}^2  \mathbf{\hat{  \mathcal{B}}_{jk}}) \psi]   \; + \;  
  2[-\frac{\hbar^2}{2m} (\mathbf{\nabla_j}\mathbf{\hat{  \mathcal{B}}_{jk}} 
  \mathbf{\nabla_j}\psi  
  \, + \, \mathbf{\nabla_k}\mathbf{\hat{  \mathcal{B}}_{jk}} \mathbf{\nabla_k}\psi)]  & \\ 
  = \;  \{\mathbf{\hat{\mathcal{B}}_{jk}}  \mathbf{\hat{H}} \psi \}  \;       
   + \;   [-\frac{\hbar^2}{2m} (\mathbf{\nabla_j}^2  \mathbf{\hat{  \mathcal{B}}_{jk}} \, + \, \mathbf{\nabla_k}^2  \mathbf{\hat{  \mathcal{B}}_{jk}}) \psi]   \; + \;  
 2[-\frac{\hbar^2}{2m} (\mathbf{\nabla_j}\mathbf{\hat{  \mathcal{B}}_{jk}} 
 \mathbf{\nabla_j}\psi  
 \, + \, \mathbf{\nabla_k}\mathbf{\hat{  \mathcal{B}}_{jk}} \mathbf{\nabla_k}\psi)].
 \end{array} 
  \end{equation} 
If we sum over the operators, $\mathbf{\hat{ \mathcal{B}}_{jk}}, $ 
this becomes:
 \begin{equation}\label{7p7} 
  \mathbf{\hat{\mathcal{B}}}  \mathbf{\hat{H}} \psi  \; 
 + \;  (-\frac{\hbar^2}{2m}) \, \{ \sum_{j<k} \, [(\mathbf{\nabla_j}^2  \mathbf{\hat{  \mathcal{B}}_{jk}} \, + \, \mathbf{\nabla_k}^2  \mathbf{\hat{  \mathcal{B}}_{jk}}) \psi   \; + \;   
 2 (\mathbf{\nabla_j}\mathbf{\hat{  \mathcal{B}}_{jk}} 
 \mathbf{\nabla_j}\psi  
 \, + \, \mathbf{\nabla_k}\mathbf{\hat{  \mathcal{B}}_{jk}} \mathbf{\nabla_k}\psi)]\}.
  \end{equation}

Substituting $\hat{\mathbf{H}}$  for $\hat{\mathbf{Q}}$ in Eq. \ref{6pa} yields:
\begin{equation}\label{7p8} 
\;     \psi^*( -\frac{1}{2} [  \mathbf{\hat{  \mathcal{B}}}
\mathbf{\hat{  \mathcal{B}}}  \mathbf{\hat{H}}  \, + \,
\hat{\mathbf{H}}  \mathbf{\hat{  \mathcal{B}}}
\mathbf{\hat{  \mathcal{B}}} ]    \;   
+ \; \mathbf{\hat{  \mathcal{B}}}  \hat{\mathbf{H}}  \mathbf{\hat{  \mathcal{B}}} \, )\psi   \, dt \; \;   + \;  \psi^* \,  \mathbf{\hat{  \mathcal{B}}}  \hat{\mathbf{H}} \, \psi  \, d\xi^* \; + \;\psi^* \,  \hat{\mathbf{H}}   
\mathbf{\hat{  \mathcal{B}}} \, \psi  \, d\xi. 
\end{equation}

With the last term in Eq. \ref{7p8} ($\;\psi^* \,  \hat{\mathbf{H}}   
\mathbf{\hat{  \mathcal{B}}} \, \psi  \, d\xi$) expanded as Eq. \ref{7p7}, after some recombination and rearrangement, the change induced by the stochastic operator, $\mathbf{\hat{  \mathcal{B}}},$ on  $  \, \psi^* \,  \hat{\mathbf{H}}   \, \psi, \, $ at each point in configuration space, expressed by Eq. \ref{7p8} results in: 
\begin{equation}\label{7p9} 
\begin{array}{ll}
 \psi^* \,    
\mathbf{\hat{\mathcal{B}}}  \mathbf{\hat{H}} \, \psi \, \, (d\xi^* \, + \, d\xi) \; 
&  \\  + \; \;
\psi^* \, (-\frac{\hbar^2}{2m}) \, \{ \sum_{j<k} \, [(\mathbf{\nabla_j}^2  \mathbf{\hat{  \mathcal{B}}_{jk}} \, + \, \mathbf{\nabla_k}^2  \mathbf{\hat{  \mathcal{B}}_{jk}}) \psi   \; + \;   
2 (\mathbf{\nabla_j}\mathbf{\hat{  \mathcal{B}}_{jk}} 
\mathbf{\nabla_j}\psi  
\, + \, \mathbf{\nabla_k}\mathbf{\hat{  \mathcal{B}}_{jk}} \mathbf{\nabla_k}\psi)]\} \, d\xi^*  
& \\  + \; \;      \psi^*( -\frac{1}{2} [  \mathbf{\hat{  \mathcal{B}}}
\mathbf{\hat{  \mathcal{B}}}  \mathbf{\hat{H}}  \, + \,
\hat{\mathbf{H}}  \mathbf{\hat{  \mathcal{B}}}
\mathbf{\hat{  \mathcal{B}}} ]    \;   
+ \; \mathbf{\hat{  \mathcal{B}}}  \hat{\mathbf{H}}  \mathbf{\hat{  \mathcal{B}}} \, )\psi   \, dt.
\end{array}
\end{equation}

The term on the top line, $ 
\, \psi^* \, \mathbf{\hat{\mathcal{B}}}  \mathbf{\hat{H}} \, \psi \, \, (d\xi^* \, + \, d\xi), \, $ displays the proportionality that we want. This shows that the \textit{apparent} increase in kinetic energy associated with the localization of the ``target" system can be accounted for by its prior interactions with other systems with which it is entangled. The rescaling of the kinetic energy in the $j$ and $k$ subsystems by the stochastic operator is correlated with a rescaling of corresponding terms in all those other subsystems. This is the key result that we wanted to show in this section.

The deviations from strict energy conservation are represented by the remaining terms, on the last two lines. It will be shown below that these occur in exactly those circumstances in which the standard nonrelativistic theory also deviates from a fully accurate accounting of energy conversion, and that they are of the same magnitude as those inaccuracies. (As already noted, these terms go to zero for stationary states and freely evolving systems.)

Because the terms on the second and third line in Eq. \ref{7p9} are of a different order in $\mathbf{\hat{ \mathcal{B}}}$ they will be examined separately. Let us begin with the  second line:
\begin{equation}\label{7pa}
\psi^* \, (-\frac{\hbar^2}{2m}) \, \{ \sum_{j<k} \, [(\mathbf{\nabla_j}^2  \mathbf{\hat{  \mathcal{B}}_{jk}} \, + \, \mathbf{\nabla_k}^2  \mathbf{\hat{  \mathcal{B}}_{jk}}) \psi   \; + \;   
2 (\mathbf{\nabla_j}\mathbf{\hat{  \mathcal{B}}_{jk}} 
\mathbf{\nabla_j}\psi  
\, + \, \mathbf{\nabla_k}\mathbf{\hat{  \mathcal{B}}_{jk}} \mathbf{\nabla_k}\psi)]\} \, d\xi^*.
\end{equation}

As shown in Section 6, the only component of $ \mathbf{\hat{\mathcal{B}}_{jk}} $ with a nonzero spatial derivative is $ \mathbf{\hat{V}}_{jk}. $ So the expression in which we are interested can be written as: 
\begin{equation}\label{7pb}
\frac{1}{(m_j+m_k)c^2}  (\Gamma_{jk})^{1/2} 
(-\frac{\hbar^2}{2m}) [\psi^* \,\psi (\mathbf{\nabla_j}^2  \mathbf{\hat{  V}_{jk}} \, + \, \mathbf{\nabla_k}^2  \mathbf{\hat{  V}_{jk}})     \; + \;   
2 \psi^* (\mathbf{\nabla_j}\mathbf{\hat{  V}_{jk}} 
\mathbf{\nabla_j}\psi  \, + \,  \mathbf{\nabla_k}\mathbf{\hat{  V}_{jk}} 
\mathbf{\nabla_k}\psi )  ]  \, d\xi^*.
\end{equation}
The effects on subsystems $j$ and $k$ represented by this expression can be analyzed separately. Consider 
$ \;  \psi^* \,\psi \mathbf{\nabla_j}^2  \mathbf{\hat{  V}_{jk}}  \; + \;   2 \psi^* (\mathbf{\nabla_j}\mathbf{\hat{  V}_{jk}} 
\mathbf{\nabla_j}\psi  ).  \;  $  The real part\footnote{Since $\xi(t)$ can be complex, multiplication by $d \xi^* $ can alter the real and imaginary parts of the expression. But, this does not affect the essential points made here and in Appendix B.} of this expression can be written as a divergence: 
\begin{equation}\label{7pc} 
 \; \; \; \;  \mathbf{\nabla_j} [ \psi^* \,\psi \, (\mathbf{\nabla_j}\mathbf{\hat{  V}_{jk}}) ] \; = \;   ( \mathbf{\nabla_j}\psi^* ) \, \psi  \, (\mathbf{\nabla_j}\mathbf{\hat{  V}_{jk}}) )  \; + \;  
\psi^* (\mathbf{\nabla_j} \psi )\, (\mathbf{\nabla_j}\mathbf{\hat{  V}_{jk}}) )  \; + \; 
 \psi^*\,\psi  (\mathbf{\nabla_j}^2\mathbf{\hat{  V}_{jk}} ) ]. 
\end{equation}
Since 
$ \; \;\; \psi^* \,\psi \, (\mathbf{\nabla_j}\mathbf{\hat{  V}_{jk}}) \; \; \;  $  vanishes at infinity, the real part integrates to zero.

In these expressions showing the effects on system $j$, the potential function,
$  \mathbf{\hat{  V}_{jk}} $, can be looked at as the action of system $k$ on system $j$. The zero result of the integration tells us that the interaction that generates the entanglement between the branch of system $k$ represented by a particular value of $ \mathbf{w_k} $ and system $j$ does not produce any net change in the real part of the energy of system $j$. Obviously, a similar conclusion holds when the roles of system $j$ and system $k$ are reversed. So within each entangled branch the real part of the expression representing the  first order effects of the stochastic process shows no violation of energy conservation.The magnitude of the local redistribution of kinetic energy within systems $j$ and $k$ can also be shown to be several orders of magnitude smaller than the changes induced by the ordinary evolution under the Hamiltonian. This is demonstrated in Appendix B.

We must still deal with the imaginary part of the expression. To do this it is helpful to compare it to a corresponding term representing the ``local"\footnote{The term 'local' in this context refers to locations in configuration space - not in physical space or spacetime.} rate of change of energy in standard quantum theory. Since in this case evolution is controlled strictly by the Hamiltonian, the question of energy conservation might seem trivial. But, even though the changes in energy sum to zero when integrated over the entire wave function 
($ \;  (d/dt)\langle\, \psi | \mathbf{\hat{H}} | \psi \, \rangle \; = \; 0 \; $), it is still useful to see how kinetic and potential energy are converted to one another at each point in configuration space, and how the local distribution of energy changes.

 The expansion of the expression representing the local changes, $ \, (d/dt)( \psi^*  \mathbf{\hat{H}}  \psi ), \; $  yields a purely kinetic term that integrates to zero separately,  
\begin{equation}\label{7pd} 
 (i\hbar^3/4m^2) [ \mathbf{\nabla}^2\psi^*  \mathbf{\nabla}^2\psi 
\, - \, \psi^*  \mathbf{\nabla}^4\psi ], 
\end{equation} 
and a term that involves a combination of potential and kinetic energy operators: 
\begin{equation}\label{7pe}   
 (i\hbar/2m) [ \mathbf{\hat{V}} ( \psi^*  \mathbf{\nabla}^2\psi 
\, - \, \psi \mathbf{\nabla}^2\psi^*) ] \; + \; 
(i\hbar/2m)[\psi^* \,\psi (\mathbf{\nabla}^2  \mathbf{\hat{  V}})     \; + \;   
2 \psi^*(\mathbf{\nabla}\mathbf{\hat{  V}} \cdot  \mathbf{\nabla}\psi)  ].
\end{equation}
This is the term in which we are interested, and it is fairly easy to interpret. The expression, 
$ (i\hbar/2m) [ ( \psi^*  \mathbf{\nabla}^2\psi \, - \, \psi \mathbf{\nabla}^2\psi^*) ]  $, from the first half of the formula can be recognized as the negative of the divergence of the probability current. In other words, it is the rate at which the probability density is changing at a given point in configuration space. The multiplication by $\mathbf{\hat{V}} $ gives the rate of change of potential energy at that point.

The second part of this formula  involving the derivatives of $\mathbf{\hat{  V}}  $ has exactly the same form as the expression for the changes induced by the stochastic collapse operator that was analyzed above, but, it is important to note that, because of the multiplication by $i$ the real and imaginary parts of the expression are interchanged. The real part in this case gives the rate at which potential energy is being converted to kinetic energy. For stationary states, there is no change at any point in either potential energy or kinetic energy (except for the imaginary term which integrates to zero). For other states there are changes in kinetic energy, implying some acceleration. As noted earlier these changes imply some small corrections to the formula for calculating kinetic energy due to relativistic increases in mass, and electromagnetic radiation (for charged particles).

It is in these situations in which some acceleration is expected  that the stochastic generalization of the standard theory generates a purely imaginary energy expression that does not necessarily integrate to zero. In Appendix B it is shown that the ratio of the magnitude of the stochastic discrepancy to the change in kinetic energy implied by the standard theory is of the same order as the correction to the kinetic energy term entailed by the  relativistic mass increase. (It is also within one or two orders of magnitude of the classical prediction of the ratio of radiated power to work done on a moving charge.) For these reasons, the discrepancy with strict conservation represented by the imaginary term should be seen as an artifact of the nonrelativistic formulation - not as an indication that the stochastic process produces an actual violation.

Now consider the terms from Eq. \ref{7p9} that are second order in the stochastic operator: 
  \begin{equation}\label{7pf} 
     \;     \psi^*( -\frac{1}{2} [  \mathbf{\hat{  \mathcal{B}}}
 \mathbf{\hat{  \mathcal{B}}}  \mathbf{\hat{H}}  \, + \,
 \hat{\mathbf{H}}  \mathbf{\hat{  \mathcal{B}}}
 \mathbf{\hat{  \mathcal{B}}} ]    \;   
 + \; \mathbf{\hat{  \mathcal{B}}}  \hat{\mathbf{H}}  \mathbf{\hat{  \mathcal{B}}} \, )\psi   \, dt.  
  \end{equation} 
The corresponding terms for momentum and angular momentum summed to zero, but in this case the fact that $\mathbf{\hat{H}} $ involves a second derivative leads to a more complicated result. With $ \mathbf{\hat{H}} \; = \;  -\frac{\hbar^2}{2m} \mathbf{\nabla}^2\, + \, \mathbf{\hat{V}} $, it is, again, clear that the terms involving $ \, \mathbf{\hat{V}} \, $ will cancel, so we need only worry about the kinetic energy expressions. Therefore, we need to evaluate:
\begin{equation}\label{7pg}   (-\frac{\hbar^2}{2m})  \psi^*[ -\frac{1}{2}  \mathbf{\hat{  \mathcal{B}}}
\mathbf{\hat{  \mathcal{B}}}  \mathbf{\nabla}^2 \psi  \,-\frac{1}{2} 
\mathbf{\nabla}^2 (  \mathbf{\hat{  \mathcal{B}}}
\mathbf{\hat{  \mathcal{B}}} \psi)    \;   
+ \; \mathbf{\hat{  \mathcal{B}}}  \mathbf{\nabla}^2\, ( \mathbf{\hat{  \mathcal{B}}}  \psi ) ]  \, dt.  
\end{equation}   
 Expanding this term we get: 
 \begin{equation}\label{7ph}  
 \begin{array}{ll}	
  (-\frac{\hbar^2}{2m})  \psi^*[ -\frac{1}{2}  \mathbf{\hat{  \mathcal{B}}}
 \mathbf{\hat{  \mathcal{B}}}  \mathbf{\nabla}^2\,\psi  \, 
 - (\mathbf{\nabla} \mathbf{\hat{  \mathcal{B}}} \cdot \mathbf{\nabla} \mathbf{\hat{  \mathcal{B}}} ) \psi 
  - \; \mathbf{\hat{  \mathcal{B}}} ( \mathbf{\nabla}^2\,  \mathbf{\hat{  \mathcal{B}}}) \, \psi \,  - \, 
  2 \mathbf{\hat{  \mathcal{B}}} ( \mathbf{\nabla} \mathbf{\hat{  \mathcal{B}}}  \cdot  {\nabla}\psi) \,
  - \, \frac{1}{2}  \,  \mathbf{\hat{  \mathcal{B}}}
  \mathbf{\hat{  \mathcal{B}}}  \mathbf{\nabla}^2 \psi   
 & \\    
\indent  \indent  \indent   \indent  \indent  \indent 
\indent \indent \indent \indent \indent 
 + \; \mathbf{\hat{  \mathcal{B}}} ( \mathbf{\nabla}^2\,  \mathbf{\hat{  \mathcal{B}}}) \, \psi \,  + \, 
2 \mathbf{\hat{  \mathcal{B}}} ( \mathbf{\nabla} \mathbf{\hat{  \mathcal{B}}}  \cdot {\nabla}\psi) \,
 + \,  \mathbf{\hat{  \mathcal{B}}}
 \mathbf{\hat{  \mathcal{B}}}  \mathbf{\nabla}^2 \psi ]  \, dt.   
\end{array}  
\end{equation} 
All of the terms cancel except for  
 $  \;  \;  \; \;  (+\frac{\hbar^2}{2m})  \psi^*   \psi  \,   
 (\mathbf{\nabla} \mathbf{\hat{  \mathcal{B}}} \cdot  \mathbf{\nabla} \mathbf{\hat{  \mathcal{B}}} )   \, dt.  $

This term clearly represents a net positive change in energy, but one must keep in mind the limitations of nonrelativistic theory in dealing with the issue of energy conservation. To understand whether this term indicates a genuine violation one should consider its relative magnitude.

As argued above, the apparent deviations from perfect conservation associated with first-order stochastic effects, which are at least four to five orders of magnitude smaller than the energy changes described by ordinary Schr\"{o}dinger evolution, are due to the limitations of a nonrelativistic formulation. The second-order discrepancy just described is more than four orders of magnitude smaller than these first-order deviations. At this level of precision these effects are well beyond the limit at which the standard nonrelativistic theory is capable of addressing issues of energy conservation.

To demonstrate this we can choose an integration time, $dt$, related to the entanglement rates, $ \, \Gamma_{jk}, \; $ as is done in Appendix B. Since the operator, $ \, \mathbf{\hat{  \mathcal{B}}},  \, $ includes the entanglement rates, $ \; \Gamma^{1/2}_{jk}, \; $ the term,  $  \;    
(\mathbf{\nabla} \mathbf{\hat{  \mathcal{B}}} \cdot  \mathbf{\nabla} \mathbf{\hat{  \mathcal{B}}} ),  \; $  includes factors of $ \; \Gamma_{jk} \; $  and 
 $ \; \Gamma^{1/2}_{jk} \Gamma^{1/2}_{jl}. \; $ Setting $dt$ equal to the inverse of the largest factor, $ \; \Gamma_{jk}, \; $ guarantees that all factors, 
 $ \; \Gamma_{jk} dt, \; $   $ \; \Gamma^{1/2}_{jk} \Gamma^{1/2}_{jl} dt, \; $ and $ \; \Gamma^{1/2}_{jk} d\xi, \; $ are less than or equal to $1$. With two factors of 
 $ \; \mathbf{\hat{  \mathcal{B}}},  \; $  and 
 $\mathbf{\hat{V}_{jk}} \, = \, \frac{e^2}{r_{jk}}, $ the second-order effects can be seen to be smaller than the first-order effects by the ratio of 
 $ \, \frac{e^2}{r_{jk}}(1/2mc^2). \, $\footnote{The form of the second order effects, 
 $  \, \psi^*   \psi  \,  (\mathbf{\nabla} \mathbf{\hat{  \mathcal{B}}} \cdot  \mathbf{\nabla} \mathbf{\hat{  \mathcal{B}}} ), \, $  is not identical to that of the first-order effects, $  \, \psi^* \, (\mathbf{\nabla} \mathbf{\hat{  \mathcal{B}}} \cdot  \mathbf{\nabla} \psi ), \, $  but, assuming there is no substantial overlap between systems, $j$ and $k$, one can see that the gradient of 
 $\mathbf{V_{jk}} \, $ must be less than $\mathbf{\nabla} \psi. $ } Using the Bohr radius as the minimum length at which a nonrelativistic description can be considered reasonably accurate, this term is less than or equal to one fourth times the square of the fine structure constant, $ \, \frac{1}{4}5.33*10^{-5} \, = \, 1.33*10^{-5}. \, $ This is far into the regime where relativistic effects must be taken into account.

 In this regime the discussion of energy conservation takes on a much different shape from that in nonrelativistic theory. Energy is no longer a scalar quantity, but, rather, a component of a four-vector. Interactions are described with the exchange of real or virtual particles, rather than with scalar potentials. In addition to mechanical energy, one must consider rest energy, relativistic mass increase, radiation,  and effects related to particle creation and annihilation. The Hamiltonian takes a substantially different form, and the fact that the number of particles is no longer constant alters the interpretation of the wave function. In relativistic theory the localization of a wave packet requires the inclusion of some anti-particle components. For all these reasons, the very small deviations from perfect energy conservation implied by the nonrelativistic formulation should not be taken as indications of a genuine violation.

To summarize, one can note that, within the standard nonrelativistic theory's range of reliability, the stochastic collapse equation conserves energy. Most importantly, it is  able to explain the apparent increase in kinetic energy associated with the localization of the wave function of the target system since the changes that it induces in that system are correlated with compensating changes in the entangled systems with which it has previously interacted. Thus, one can conclude that the collapse equation guarantees conservation of energy to the extent that one can reasonably expect for a nonrelativistic formulation.

\section{Some Further Examples}
\label{sec:8}

The formal arguments of the last few sections show that the stochastic collapse equation of Section 4 strictly conserves momentum, orbital angular momentum, and energy (with the nonrelativistic limitations discussed above). However, without some serious simplifications it is not feasible to formally analyze 
the interactions of elementary and macroscopic systems. Nevertheless, it is possible to describe in qualitative terms how the relevant quantities are conserved in these situations. In this section I will present a few examples to illustrate this point.

We can begin with a simple example involving momentum.  Consider a particle of mass, $m$, with a Gaussian wave function that is centered at the origin ($x_i = 0$) with an initial width of $2a$, and momentum, zero, in the sense in which conserved quantities were defined in Section 1:  
\begin{equation}\label{8p1}
\int \, dx\; \psi^*(x)(-i\hbar(\partial{\psi(x)}/\partial{x})) \; = \; 0.
\end{equation}  
The wave function corresponding to this probability distribution is  
$ \psi_i(x) = N_i *e^{-{\frac{x^2}{ 4a^2 }} }$. $N_i$ is a normalization factor. After a time, $t$, the wave function will have expanded:  
\begin{equation}\label{8p2}
 \psi_t(x) = N_t *e^{-{\frac{x^2}{ 4(a^2+(it\hbar/2m) }} }.
\end{equation} 
If the particle is detected at time, $t$, in a localized region centered at $x_f$ between $x_f-\epsilon$ and $x_f+\epsilon$, the net momentum contributed by the pre-detection wave function can be calculated by applying the momentum operator to the portion of $\psi_t$ that lies within this region, and then normalizing: 
\begin{equation}\label{8p3} 
\begin{array}{ll}
[\int_{x_f-\epsilon}^{x_f+\epsilon} \, dx\; \psi_t^*(x)(-i\hbar(\partial{\psi_t(x)}/\partial{x})) ] \; 	/ \; [\int_{x_f-\epsilon}^{x_f+\epsilon} \, dx \; \psi_t^*(x)\psi_t(x)] 
& \\   \;\;\;\;\;\;\;\; 
 \approx \, 
i\hbar \frac{x_f }{ 2(a^2+(it\hbar/2m) }  \, = \, 
\frac{mx_f}{t-2ima^2/\hbar}. 
\end{array}
\end{equation} 
This expression accords (roughly) with our intuitive expectation for classical systems that a particle of mass, $m$, traveling from the origin to a point, $x_f$, in time, $t$, will carry momentum approximately equal to $mx_f / t $.

Another way of deriving this expression is to expand the wave function at time, t, in the momentum basis: 
\begin{equation}\label{8p4}
 \psi_t(x) = \int dp \; \psi(p) *e^{i[(px/\hbar)-(p^2t/(2m\hbar))] },
\end{equation} 
 and then determine   which momentum eigenstates dominate the wave function at the point, $x$. In the momentum basis the wave function can be represented as,  
 \newline 
$ \psi_(p) = N_p *e^{-{\frac{p^2a^2}{ \hbar^2 }} }$. Inserting this expression into the integral we get: 
\begin{equation}\label{8p5} 
 \psi_t(x) = C*\int dp \; e^{-(a^2/ \hbar^2 
	+ it/(2m\hbar))p^2 - (ix/\hbar)p },
\end{equation} 
where $C$ is a constant. We can express the integrand as a Gaussian by making the substitution, $ p\, \rightarrow \, p'$, and completing the square of the exponent: 
\begin{equation}\label{8p6}
p' = (\sqrt{(2ma^2+it\hbar)/(2m\hbar^2)})p \, - \, (ix/2\hbar)(\sqrt{(2m\hbar^2)/(2ma^2+it\hbar)}). 
\end{equation} 
Note that $p'$ depends on the value of $x$.\footnote{For convenience, $p'$ has been defined as a dimensionless variable (unlike $p$ which has the dimensions of momentum).}  At any particular value of $x$ the wave function will be dominated by momentum terms centered at $p'(x) = 0$. Setting $ p'(x) \, = \, p'(x_f) \, = \, 0$, and solving for $p$ gives $ p \, = \, \frac{mx_f}{t-2ima^2/\hbar} $, the result derived above.

The question that must be addressed regarding conservation of momentum is how a particle with zero initial momentum can be  subsequently detected with momentum, $mx_f / t $.\footnote{Note that we are \textit{not} interested in the particle's \textit{post-detection} wave function, which includes a  range of momenta of 
	$\hbar / (2\epsilon)$, resulting from the interaction with the \textit{measurement} apparatus. The quantity of interest is the momentum transferred \textit{from} the particle \textit{to} the apparatus.} 
The apparent paradox arises because we simply stipulated that the particle to be measured could be described by a Gaussian wave packet with width, $2a$. This wave packet evolves to yield a probability distribution at time, $t$, of 
\begin{equation}\label{8p7}
 \psi_t(x)^**\psi_t(x)  = 
N_t^2 *e^{-{\frac{x^2}{ 2(a^2+(t^2\hbar^2/(4m^2a^2))) }} }. 
\end{equation}
The  standard deviation associated with this expression is $ [a^2+(t^2\hbar^2/(4m^2a^2))]^{\frac{1}{2}} $. For elementary particles initially localized in small regions the width of the wave function expands extremely rapidly. For a free electron (mass $ \approx 10^{-30} kg$) it  would stretch from atomic dimensions to about one meter in a microsecond. It is clear that highly localized wave functions of free particles typically will not occur without interacting with some large system that acts to prepare the particle in the assumed state. This large system can be either a preparation apparatus in a laboratory or a naturally occurring system consisting of a very large number of particles. What has been left out of the analysis is any consideration of the possible entanglement between the particle and the system that placed it in the ``initial" state.

As discussed earlier the role of the preparation apparatus is typically ignored because the effect on the state of the apparatus by the system to be measured is extremely small, and so the entanglement between them is also quite small. However, while this entanglement is small it is not zero. Whenever two systems interact they have some effect on one another.  The demonstrations in \cite{Gemmer_Mahler}, \cite{Durt_a}, and \cite{Durt_1} make it clear that entanglement is a \textit{generic} result of interactions between subsystems. In particular, any set of interactions that results in a particle being subject to clearly distinguishable possible measurement outcomes (such as a range of different position states)  generates some degree of entanglement. When a subsequent measurement induces a collapse of this subsystem to one of those component states, it also collapses the state of the (often much larger) subsystem that was involved in the prior set of  interactions. The resulting state of the preparation apparatus is one that reflects only those interactions and exchanges of momentum that are consistent with what is eventually detected by the measurement apparatus.

To see how this example fits with the formal discussions in previous sections, note that the portion of the initial wave function that is detected in the region between $x_f-\epsilon$ and $x_f+\epsilon$ represents the branch of wave function to which the state is collapsed by the measurement. It is entangled with a corresponding branch of the wave function of the preparation apparatus. The normalization factor that was used in the initial derivation of the result, 
$ p \, = \, \frac{mx_f}{t-2ima^2/\hbar}, $ is just the pre-measurement squared amplitude of that branch. The appearance of an imaginary component in the calculated momentum is an indication that the state of the particle is entangled with the that of the apparatus; it occurs because the integration of the momentum operator was done over only a portion of the entangled wave function.

The next example is based on a very interesting article by Aharonov, Popescu, and Rohrlich (APR)\cite{APR}. In the article the authors present an especially striking example of the apparent violation of energy conservation. They describe how a particle that is initially in a superposition of strictly low-energy states can be detected in a high-energy state. This superposition generates superoscillations. These special states are characterized by the fact that, over limited intervals, they can oscillate arbitrarily faster than the fastest Fourier component. The authors describe the  situation as a low-energy particle that looks like a high-energy particle in the center of the box. Initially, the box is completely closed, but an opening mechanism slides along the top of the box, and inserts a mirror for a very short time, possibly allowing high-frequency segments of the wave function in the exposed region to escape. This action results in some (very small) probability that the particle can be detected in a high-energy state outside the box. The detection of an initially low-energy particle in a high-energy state constitutes the apparent paradox. The example is carefully constructed to eliminate obvious sources for the extra energy. The energy of the opener is essentially unchanged, and correlations between the opener and particle are minimized. The opening mechanism is the only system in the example other than the particle that is treated as a quantum system. Since it does not supply the extra energy, and since the total Hamiltonian is time-independent, APR pose the question of where the particle's energy comes from.

The question of where the extra energy comes from is particularly puzzling because the particle is detected in a high energy eigenstate that was totally absent from the initial wave function. This is what distinguishes this case from the simpler example of a spreading Gaussian wave packet discussed earlier. The very unusual feature of the Hamiltonian that makes this possible is that it allows the opener to act on the external potential. In other words, it allows a quantum subsystem to act on a classical system - the box.  

Of course, the action of one system on another implies that there must be an interaction between them. This interaction leads to some entanglement, and this is why the box must be treated as a quantum system. It is also true that, although the entanglement between the particle and the opening mechanism mirror is minimal, it does play a role in the evolution of the system. The entanglement chain enables the opener to serve as an intermediary, transferring energy from the box (and the preparation apparatus) to the particle.  The ``extra" energy supplied to the particle comes at the expense of the box and the preparation apparatus. The very slight changes in the states of these macroscopic systems that result from the collapse of the overall wave function are completely adequate to account for a substantial increase in energy by the elementary system.

The ingenious way in which the APR example is constructed helps to bring into sharp focus the kinds of apparent paradox that can arise in these situations. However, even though the potential for paradox is often hidden, it is a \textit{generic} feature of situations that are described with a combination of classical and quantum systems. This was pointed out by Gemmer and Mahler\cite{Gemmer_Mahler}. They emphasize that a full description of these types of cases would recognize the entanglement between the microscopic and macroscopic systems: \begin{quote} ``Thus it is, strictly speaking, unjustified to describe a particle in a box, \textit{which is part of an interacting quantum system}, by a wave-function" (italics added).\end{quote}

Of course, such combined quantum-classical descriptions are extremely useful for several purposes. Since the amount of entanglement between the microscopic and macroscopic systems is so small, key aspects of the situation can be presented more clearly. In addition, the mathematical treatment can be greatly simplified by the use of \textit{factorizable approximations}. Both Gemmer and Mahler\cite{Gemmer_Mahler}, and Durt\cite{Durt_a,Durt_1} examine these approximations in some detail. In \cite{Gemmer_Mahler} the authors develop an estimate for the error involved in using the product state description. They derive an expression for the inner product between the approximate and exact quantum states and relate it to the purity, $P$, where their entanglement measure is defined as $1-P$.\footnote{P= Tr($\rho^2)$ where $\rho$ is the \textit{reduced} density matrix.} Their relationship between the inner product and entanglement is generally similar to the example calculation given in Appendix A, although their entanglement measure is somewhat  different from the relative von Neumann entropy.

The relationship between entanglement and factorizability has been examined from a somewhat different perspective by Dug\'{i}c, Jekn\'{i}c-Dug\'{i}c, and  Arsenijev\'{i}c,\cite{Dugic_1,Dugic_2}, and also by Thirring, Bertlmann, K\"{o}hler, and Narnhofer\cite{Bertlmann_2,Bertlmann_1}. These authors emphasize the point that entanglement depends on the way in which a composite system is decomposed into subsystems. In \cite{Bertlmann_2} it is shown that (pure) entangled states can always be transformed into factorizable states by a unitary transformation that alters the tensor product structure of the Hilbert space. Ordinarily we construct the Hilbert space for a combined system by forming the tensor product of Hilbert spaces corresponding to the individual subsystems. In the photon/beam-splitter example of Section 3 one sector would correspond to the photon and the other to the beam-splitter. However, in order to analyze the interaction between systems it is usually more convenient to redefine the variables involved in order to decouple the equations. This decoupling implies a switch to a different tensor product structure.

For example, consider a system of two particles interacting through a distance-dependent potential. In these situations the wave functions of the particles almost certainly become entangled, but the factorizable description in which one wave function corresponds to the center of mass, and the other to the reduced mass does not reflect the entanglement between the interacting systems. The redefinition of the coordinates as $R_{cm} = \frac{m_1r_1+m_2r_2}{(m_1+m_2)} $ and $ r_{rel} = r_2-r_1 $ is a rotation of the coordinates in configuration space. This rotation is associated with a unitary transformation in the Hilbert space that changes the tensor product structure. When one of the systems is macroscopic and the other microscopic one has $ m_1 \gg m_2$. This implies that $M_{tot} \approx m_1$, $\mu_{reduced} = \frac{m_1*m_2}{m_1+m_2} \approx m_2$, and $R_{cm} \approx r_1$. If the center of mass is placed at the origin then we also have  $r_{rel} \approx r_2 $. Thus, it is very tempting to make the identifications suggested by the approximate equivalence, and to ignore the very small entanglement between the interacting systems. But we must remember that this is an approximation based on a deformation of the original tensor product structure. In any accounting of conserved quantities of the individual subsystems, we must keep track of possible exchanges brought about by the interaction between them. This means that we must recognize the residual entanglement induced by that interaction.

This reinforces the central role that interactions play in determining the way in which the wave function should be decomposed when addressing the issue of conservation. Both the tensor product structure and the basis of the wave function should be selected by reference to the interactions responsible for the exchange of the relevant quantities.

The final example concerns the conservation of angular momentum in a situation involving both spin and magnetic fields. Since these are fundamentally relativistic in character it is not possible to describe them in nonrelativistic theory using only conservative interaction potentials. Hence, this case is not covered by the formal proofs of the previous sections. Nevertheless, the qualitative account presented here does suggest that entanglement relations can also help maintain conservation laws in a relativistic setting.

The original Stern-Gerlach experiment\cite{G_S_1,G_S_2,G_S_3} used spin-${\frac{1}{2}}$ particles with \textit{unknown} initial states. However, before considering this more general situation it will be easier to begin the discussion by assuming that the particles have been previously placed into an x-up state (parallel to the direction of motion). The inhomogeneous magnetic field of the S-G apparatus is in the z-direction and separates  z-up and z-down components of the x-up particles. Downstream detectors then measure the z-spin state, collapsing it to one of the two possibilities. Since the collapse seems to eliminate any information about the phase relationship between the z-spin branches, it might not be obvious how the x-up angular momentum could be conserved.

The apparent elimination of relative phase information comes about because descriptions of Stern-Gerlach experiments typically treat the apparatus and magnetic field as classical entities. The Pauli equation describes the spin-${\frac{1}{2}}$ particle with a two-component wave function that is \textit{acted on} by the magnetic field of the apparatus. This action is represented by the relevant term in the Hamiltonian: 
$ - \frac{q\hbar}{2mc}\mathbf{\sigma} \cdot \mathbf{B}
\,\left[\begin{array}{c}
\phi_1  \\
\phi_2 \\
\end{array}\right].$
In this representation the (external) magnetic field acts on the two components of the particle wave function, but these components do not act on the apparatus. Thus, any information about the relative phase differences that distinguish an x-up state from an x-down state appears to be lost when one of the two z-spin branches is subsequently detected.

The approximate, semi-classical description of the action treats the S-G apparatus as simply separating the z-up and z-down components of the particle state. In a more complete description the wave function would include both the particle and the apparatus as quantum systems, and the Hamiltonian would reflect the \textit{interaction} between these two systems. A sufficiently detailed account of the interaction would show that the effect of the particle on the apparatus is different depending on whether it is initially in an x-up or x-down state.

The crucial point is that the the magnetic effects of the z-up and z-down components on the state of the apparatus will differ (very slightly) depending on their relative phases (corresponding to an initially x-up or x-down state).\footnote{The absence of, say, a z-down component would also result in a different apparatus state.}  According to \cite{Gemmer_Mahler}, \cite{Durt_a}, and \cite{Durt_1}, subsequent to the interaction between the apparatus and particle there must be some entanglement between them. This can be represented as \newline 
$(1/\sqrt 2)(|\mathcal{P}_{up}\rangle|z-up\rangle + |\mathcal{P}_{down}\rangle|z-down\rangle )$, where $\mathcal{P}$ represents the (preparation) S-G apparatus.

The resulting combined state will vary depending on the initial phase relationship of the z-spin components, and, hence, must contain the initial phase information. (Note also that, since at this stage prior to any collapse the evolution is unitary, the x-angular momentum of the combined system must be conserved.) This means that as the $|z-up\rangle$ and $|z-down\rangle$ branches are deflected upward or downward, either some rotational motion about the x-axis must be induced, or the x-angular momentum must be transmitted to the apparatus. 

Although the magnetic field is asymmetric the interactions between the $|z-up\rangle$ and $|z-down\rangle $ branches and the apparatus are generally similar. Therefore, the information about the initial x-angular momentum is shared, approximately equally, between the two component states, $|\mathcal{P}_{up}\rangle|z-up\rangle$ and 
$|\mathcal{P}_{down}\rangle|z-down\rangle $. So these two correlates reflect the conversion of an initially x-up (or x-down) state to either a z-up or z-down state. Roughly speaking, for the case of an initially x-up particle, they would be equivalent to the rotation of the x-up particle to a z-up state by a clockwise rotation about the y-axis (or some equivalent set of rotations) and a counterclockwise rotation to a z-down state. For an initially x-down state these rotations would be reversed. (If it had been in a z-up state there would be a simple deflection.) So, in the case of an initially x-up state that is subsequently detected in a z-up state the \textit{apparent} creation of z-up angular momentum is actually supplied by the torque that the apparatus exerts on the spin state of the particle, and the apparent elimination of x-up angular momentum  is, in reality, a transfer of angular momentum from the particle to the apparatus (or to a rotational motion about the x-axis).

This argument can be extended to cases in which the state of the system to be  measured is unknown prior to its interaction with the preparation apparatus. The assumption of a specific phase relationship between the z-up and z-down components of the spin-${\frac{1}{2}}$ particle was convenient for the purposes of illustration,  but it was not really essential. If one assigned the components arbitrary amplitudes of $\alpha$ and $\beta$ then the superposed effects of the particle components on the preparation apparatus would result in superposed apparatus states corresponding to appropriate rotations of the initial particle spin state to z-up and z-down states. Angular momentum would again be conserved.

To avoid confusion over the fact that the x-angular momentum and the z-angular momentum do not commute, recall from Section 1 that the way in which these quantities are being defined is as in Eq. \ref{1p2} ($ q \; = \;\langle\psi |\textbf{ Q}|\psi\rangle$), where $\psi$ is taken to represent the total system. The noncommutativity of two quantities simply prevents a system from simultaneously being in an eigenstate of both. The more general definition of these quantities can be applied to any pure state.

 \section{Discussion}
 \label{sec:9}

Given the central role of conservation laws in physical theory one would like to see that they hold, without qualification, in all circumstances. The present argument for this possibility is based on the assumption that the apparent collapse of the wave function associated with quantum measurements is an actual physical occurrence that is induced by the interactions that generate entanglement. The focus on entangling interactions is motivated, in part, by the recognition of the role that entanglement plays in mediating collapse and the way in which it can affect conserved quantities. The interactions through which these quantities are exchanged and by which entanglement is generated define a natural decomposition of the wave function, and thus determine the basis into which collapse occurs.

The work of Gemmer and Mahler\cite{Gemmer_Mahler} and Durt\cite{Durt_a,Durt_1} showing that entanglement is a generic result of interactions between systems also suggests that some small amount of entanglement almost always survives preparation and measurement procedures. This means that no matter how we define the boundaries of a system for the purposes of analysis or experimentation, that system is almost always entangled with its environment. Therefore, to address the issue of conservation we must free ourselves of the received notion that conserved quantities are well-defined only in those situations in which the system under consideration happens to be in an eigenstate of the relevant observable. It is necessary to adopt the definition that is employed in the derivation of conservation laws: $ \mathbf{q} \; = \;\langle\psi|\hat{\mathbf{Q}}|\psi\rangle$, where $\psi$ includes all entangled subsystems, both microscopic and macroscopic.

To formalize the hypothesis that entangling interactions induce wave function collapse one can build on the substantial body of work dealing with stochastic collapse equations. In this approach the unitary Schr\"{o}dinger equation is supplemented with a stochastic operator designed to bring about collapse to a particular basis. To pick out the natural decomposition of the wave function defined by entangling interactions the stochastic collapse operator can be constructed from the potential energy functions that appear in the Schr\"{o}dinger equation. These potentials must be conservative and interactive. The operator must also reflect the amount of entanglement generated by an interaction and the rate at which it is generated. These requirements have dictated the form of the collapse equation described in Section 4. They have also essentially determined the value of the equation's parameters, eliminating the need to introduce new physical constants.

It is worth noting that, although the argument presented here is based on the possibility of insuring strict conservation in individual cases, this form of the collapse equation was originally proposed as a way to reconcile the nonlocal nature of wave function collapse and relativity\cite{Gillis_2}.\footnote{The version of the collapse equation presented in that work was identical to that described here, except that the timing parameter was left unspecified.} The approach taken in that earlier work was to identify the critical feature that is common to relativity and quantum theory, namely, the prohibition of superluminal information transmission. To exploit that feature it was necessary to explain the transmission of information in terms of fundamental physical processes. Since information is transmitted by the interactions that establish correlations between physical systems this led to the hypothesis that it is these entangling interactions that induce collapse.\footnote{The preferred reference frame implicit in the nonrelativistic formulation presented here should be seen simply as a special case of a randomly evolving spacelike hypersurface as argued in the work cited.}

It has often been noted that the probabilistic character of wave function collapse is the key to reconciling the nonlocality of entanglement with relativity. The argument presented here suggests a new perspective on the relationship between probability and entanglement. One might say that, given the probabilistic character of some fundamental physical processes, entanglement is nature's way of enforcing conservation laws.

\section{Acknowledgments}

I would like to thank Miroljub Dug\'{i}c for bringing some additional references to my attention.

\appendix
 \section{Preparation Entanglement Measure }

The entangled state used as an example in section 3 involved a photon, $|\gamma\rangle$, and a beam-splitting mirror, $|b \rangle$. A formula for the photon/beam-splitter entanglement is derived here in order to illustrate that, although it is very small, the entanglement is not zero. For these simple cases the relative von Neumann entropy\cite{von_Neumann} serves as a useful measure of entanglement. It is defined as:  \begin{equation}\label{ap1}  
 S(\rho_{\gamma}) \, \equiv \, -Tr[\rho_{\gamma} \log(\rho_{\gamma} )]
  \, = \,  S(\rho_b)    \, \equiv \, -Tr[\rho_b \log(\rho_b)], 
 \end{equation}  
  where $\rho_{\gamma}$ and $\rho_b$ are the reduced density matrices. If the two branches of the photon wave function have equal amplitudes, then the entangled state that results when one of the branches is reflected by the mirror and the other is transmitted can be represented as  
   $(1/\sqrt 2)(|\gamma_r\rangle|b_r\rangle + |\gamma_t\rangle|b_t\rangle ).$ Since  $|b_r\rangle$ and $ |b_t\rangle $ are not orthogonal, to compute the entropy one first does a Schmidt decomposition of the state into the form,   $\mu|\gamma_m\rangle|b_m\rangle \,+ \,\nu|\gamma_n\rangle|b_n\rangle,$ where both the pairs, 
  $|\gamma_m\rangle$ and $ |\gamma_n\rangle$, and  
   $|b_m\rangle $ and $|b_n\rangle,$ are orthogonal. To simplify notation, let $ \mu \, = \, |\mu|, \, $ and $ \nu \, = \, |\nu|. \, $
  The entropy is then given by 
   \begin{equation}\label{ap2}  
   -\mu^2\log{(\mu^2)} - \nu^2\log{(\nu^2)}.    
 \end{equation}

 To perform the decomposition, we first represent the inner product between $|b_r\rangle$ and $ |b_t\rangle $  as 
  \begin{equation}\label{ap3} 
|\langle\;b_r\; | b_t\; \rangle | \; = \eta  \; = \; 1-\delta. 
 \end{equation} 
Intuitively, one expects that $ \, \delta\, \ll \, 1. $  In a slightly more formal way it can be noted that, in order to get a fairly ``clean" reflection of the photon, the position of the mirror must have an uncertainty much less than $\lambda$, where $\lambda$ is the wavelength of the photon. Hence, the representation of the beam-splitter state in momentum space prior to reflection, 
$| b_0\; \rangle,$ must have a spread much greater than $ h / \lambda $ where  $ h / \lambda $ is the momentum of the photon. So the change in the mirror state resulting from the exchange of momentum with the photon must be very small.

The assumption that the amplitudes of the two photon branches are equal implies a symmetrical form for the orthogonal states. (In general, calculating the Schmidt decomposition is more complicated.) The unnormalized states are:
 \begin{equation}\label{ap4} 
\begin{array}{ll}
 \; |\gamma_m'\rangle \;\; = \;\; \;\;\;  |\gamma_r\rangle + |\gamma_t\rangle \;  ; \;\;\;  
\;\;\;  \;  \;\;\;  \;\;\;  \;\;\;    \;\;\;    \;
|\gamma_n'\rangle \;\; = \;\;\; \;\;    |\gamma_r\rangle - |\gamma_t\rangle \;  ;  
 & \\
 \; |b_m'\rangle \;\,  = \;\;\;\;\, |b_r\, \rangle  + |b_t \, \rangle ; \;\;\;   \;\;\;   \;\;\; \;\; \; \;
\;\;\;   \;\;\;  \;\;\;  |b_n'\rangle \; = \, \;\;\;\;  |b_r\, \rangle  - |b_t \, \rangle. 
\end{array}
 \end{equation}
Using the inner product, Eq. \ref{ap3} ($|\langle\;b_r\; | b_t\; \rangle | \; = \eta $), to normalize the states we get:  
 \begin{equation}\label{ap5}  
\begin{array}{ll}
 \; |\gamma_m\rangle \;\; = \;\; \;\;\;  \frac{1}{\sqrt{2}}\;\; [ \;|\gamma_r\rangle + |\gamma_t\rangle \; ] ; \;\;\;  
\;\;\;  \;  \;\;\;  \;\;\;  \;\;\;    \;\;\;    \;
|\gamma_n\rangle \;\; = \;\;\; \;\;\frac{1}{\sqrt{2}}\;\;[ \; |\gamma_r\rangle - |\gamma_t\rangle \; ] ;  
 & \\
 \; |b_m\rangle = \, \frac{1}{\sqrt{2(1+\eta)}}[|b_r\, \rangle  + |b_t \, \rangle ] ; \;\;\;   \;\;\;   \;\;\; \;\;
\;\;\;   \;\;\;  \;\;\;  \;\; \, |b_n\rangle \; =  \;\;  \frac{1}{\sqrt{2(1-\eta)}}[|b_r\, \rangle  - |b_t \, \rangle ]. 
\end{array}
 \end{equation}
Rearranging and substituting into the original form for the entangled photon-mirror state, we get: 
 \begin{equation}\label{ap6}
 (1/\sqrt 2)(|\gamma_r\rangle|b_r\rangle + |\gamma_t\rangle|b_t\rangle ) \; = \;  \sqrt{\frac{(1+\eta)}{2}}(|\gamma_m\rangle|b_m\rangle) \;  
+\; \sqrt{\frac{(1-\eta)}{2}}(|\gamma_n\rangle|b_n\rangle).    
 \end{equation}
So the coefficients in the Schmidt decomposition are:  
 \begin{equation}\label{ap7}
\mu \;  =  \; \sqrt{\frac{(1+\eta)}{2}} \; = \; \sqrt{1 - \frac{\delta}{2}}; \; \; \; \; \;
\nu \; = \; \sqrt{\frac{(1-\eta)}{2}} \; = \; \sqrt{\frac{\delta}{2}}.     
 \end{equation}
Thus, the entanglement measure given by the relative entropy, Eq. (\ref{ap2}),
is 
 \begin{equation}\label{ap8}
 - [ (1 - \frac{\delta}{2})\log(1 - \frac{\delta}{2})  
+ (\frac{\delta}{2})\log( \frac{\delta}{2}) ].   
 \end{equation}
For  $ \delta \, \ll \, 1 $, we can approximate by: 
 \begin{equation}\label{ap9}  
 - [ 1 *(- \frac{\delta}{2})  + (\frac{\delta}{2})*\log( \frac{\delta}{2}) ] \; = \;(\delta/2)[1-\log(\delta/2) ],  
 \end{equation}
the form given in section 3.

\section{Comparing Magnitudes of Stochastic and Hamiltonian Effects }

The apparent discrepancies with strict energy conservation predicted by the first-order stochastic terms are several orders of magnitude smaller than the rate at which potential energy is converted into kinetic energy according to the Schr\"{o}dinger equation. Changes in kinetic energy lead to deviations from the nonrelativistic formula for kinetic energy due to relativistic changes in inertial mass. At low speeds the ratio of this deviation to the rate of change of kinetic energy is between $10^{-4}$ and $10^{-5}$. I will show here that the ratio of the stochastic discrepancies to kinetic energy conversion is approximately the same. Since these discrepancies occur at the same level as relativistic deviations and in exactly the same situations they cannot be taken as evidence of genuine violations of energy conservation. Both the relativistic deviations and the stochastic discrepancies should be seen as resulting from the failure of nonrelativistic quantum theory to describe certain forms of energy.

In standard, nonrelativistic quantum theory the rate at which potential energy is converted into kinetic energy at a location in configuration space (as described in Section 7) is: 
\begin{equation}\label{bp1}    
 (i\hbar/2m)[\psi^* \,\psi (\mathbf{\nabla_j}^2  \mathbf{\hat{V}_{jk}})     \; + \;   
2 (\mathbf{\nabla_j}\mathbf{\hat{V}_{jk}} \cdot  \mathbf{\nabla_j}\psi)  ]. \, 
\end{equation}
The corresponding term in the collapse equation is: 
\begin{equation}\label{bp2}     
 (-\frac{\hbar^2}{2m})  [\psi^* \,\psi (\mathbf{\nabla_j}^2  
\mathbf{\hat{  V}_{jk}}  \; + \; 2 \psi^* (\mathbf{\nabla_j}\mathbf{\hat{  V}_{jk}} 
\mathbf{\nabla_j}\psi    ] \frac{1}{(m_j+m_k)c^2} (\Gamma_{jk})^{1/2}  \, d\xi^*.   
\end{equation} 
Because of the multiplication of the Schr\"{o}dinger term by $i$ the real and imaginary parts of the expressions are interchanged.\footnote{Aside from the multiplication by $d \xi^*$.} It is the imaginary part of the stochastic term that does not necessarily integrate to zero; so, our primary interest is to compare it to the real part of the Schr\"{o}dinger term. The form of the two expressions is identical, although there is a dimensional difference between them. In order to determine the ratio of the effects, the Schr\"{o}dinger term can be integrated over a small time interval. The natural interval to choose is the minimum time in which an initially unentangled, nonrelativistic, two-particle system can reach full entanglement. To determine this time it is necessary to develop an expression for the entanglement rate, $ \Gamma_{jk} $.

As discussed in Appendix A, the entanglement generated by the interaction is related to the norms, $\mu$ and $\nu$, of the orthogonal branches by Eq. \ref{ap2} 
\newline 
($ -\mu^2\log{(\mu^2)} - \nu^2\log{(\nu^2)}.$) The rate of change of entanglement is, therefore:
\begin{equation}\label{bp3} 
\begin{array}{ll} 
\;\;\;\;
 - \frac{d}{d \mu}\, {\big [} \, \mu^2\log{(\mu^2)}\, {\big ]} \, {\big (}\frac{d \mu}{dt} \, {\big )}
- \frac{d}{d \nu}\, {\big [} \, \nu^2\log{(\nu^2)} \, {\big ]} \, {\big ]} \, {\big (}\frac{d \nu}{dt} \, {\big )}	
&  \\ 
= \; \; 
- \, {\big [} \,  2 \mu \log{(\mu^2)} 
\, + \, \frac{2\mu^2} {\mu}  \, {\big ]} \, {\big (}\frac{d \mu}{dt} \, {\big )}  \; \; 
-  \, {\big [} \,  2 \nu \log{(\nu^2)} 
\, + \, \frac{2\nu^2} {\nu} \,  {\big ]} \, {\big (}\frac{d \nu}{dt} \, {\big )}.
\end{array}
\end{equation}

Since $ \mu^2 \; = \; 1 \, - \,  \nu^2, \,  $  we get 
\begin{equation}\label{bp4} 
 \, 2\mu{\big (}\frac{d \mu}{dt} \, {\big )}  \,  =  
\, - \, 2\nu{\big (}\frac{d \nu}{dt} \, {\big )}  \;  \;  
   \Longrightarrow
 \; \; {\big (}\frac{d \mu}{dt} \, {\big )}  
\,  =  \, - \, \frac{\nu}{\mu}{\big (}\frac{d \nu}{dt} \, {\big )}.
\end{equation}
Thus, the entanglement rate is given by:  
\begin{equation}\label{bp5}    	
  \; \;  2{\big (}\frac{d \nu}{dt}  {\big )}   
{\Big [} \, \nu \log(\mu^2)  
\, + \, \nu  \,  -  \,  \nu \log{(\nu^2 )}
\, - \, \nu \,  {\Big ]} \;     
=  \; \,   2 \nu {\big (}\frac{d \nu}{dt}  {\big )}    
{\Big [} \, \log(\mu^2)  
\,  -  \, \log{(\nu^2)} \,  {\Big ]} . 
\end{equation}
(We are assuming that $\nu$ represents the norm of the smaller of the two orthogonal components.)

In order to determine $ \frac{d \nu}{dt}  $ we need to calculate the rate at which  the wave function is changing as a result of the interaction.  In a small increment of time, $dt$, the change induced in the wave function by an interaction between two elementary subsystems, 
$ \phi_j$ and $\chi_k$, is 
\begin{equation}\label{bp6} 
 \; d\psi_{jk} \; = \; -\frac{i}{\hbar} \mathbf{\hat{V}_{jk}}\phi_j \chi_k \; dt. 
\end{equation}
The interaction will generate components of the wave function that are orthogonal to what the system would have evolved to if there were no interaction:\footnote{In a stationary state the orthogonal components have norm, zero. This is why stationary states remain stationary and no additional entanglement is generated.}  
\begin{equation}\label{bp7} 
 \, -\frac{i}{\hbar} (\mathbf{\hat{V}_{jk}} -  \langle \psi_{jk} |\mathbf{\hat{V}_{jk}}|\psi_{jk} \rangle) \, \psi_{jk} \, dt.
\end{equation} 
The norm of the rate of change of this vector can be designated as $\epsilon$. It is: 
\begin{equation}\label{bp8}  
  \;  \; \parallel  - \, \frac{i}{\hbar} (\mathbf{\hat{V}_{jk}} -  \langle \psi_{jk}
 |\mathbf{\hat{V}_{jk}}|\psi_{jk} \rangle) \, \psi_{jk}    \parallel  \; 
=  \;   \frac{1}{\hbar} \sqrt{(\langle \psi_{jk} |\mathbf{\hat{V}_{jk}}^2|\psi_{jk} \rangle) -  \langle \psi_{jk} |\mathbf{\hat{V}_{jk}}|\psi_{jk} \rangle) \langle \psi_{jk} |\mathbf{\hat{V}_{jk}}|\psi_{jk} \rangle)}.  \;  
\end{equation}

For two initially unentangled systems we have 
$ \mu \, = \, 1;  \; \; \nu \, = \, 0.  $ Given the orthogonal relationship between the components, and that $ \mu^2 \, + \, \nu^2 \; = \; 1,  $ one can see that 
$ \nu \; = \; \sin (\epsilon t). $ The interacting systems reach maximum entanglement when $  \nu^2 \; = \; \frac{1}{2}.  $  This occurs when $ \epsilon t \; = \; \frac{\pi}{4}.   $  So the entanglement rate is 
$ \,  4 \epsilon / \pi. \,$ The entanglement time is the inverse of this quantity: $ \, t_E \; = \; (\pi / 4 \epsilon) . $

The specific value of $\epsilon$ represented by the expression above will depend on the exact shape of the wave functions and the separation between the systems, $j$ and $k$. A conservative maximum for this value is $ \frac{1}{2}  (\mathbf{\hat{V}_{jk}} / \hbar). $ For two elementary charged particles the electrostatic potential is $ e^2 / r. $ To stay within the limits in which nonrelativistic theory can be applied with reasonable accuracy we can set a minimum separation equal to the Bohr radius and consider the interaction of two electrons. With $ r \; = \; r_B $ we get an entanglement time, $ t_E, $ equal to 
\newline 
$ \, t_E \; = \; \pi \hbar / (4 \epsilon  ) \;  
= \; r_B \pi \hbar / (2 e^2  ) \; =\; 3.81*10^{-17} \, $  seconds.

We can now compare the magnitudes of the first-order stochastic discrepancy, Eq. \ref{bp2},   
$ \;\;\;  (-\frac{\hbar^2}{2m})  [\psi^* \,\psi (\mathbf{\nabla_j}^2  
\mathbf{\hat{  V}_{jk}}  \; + \; 2 \psi^* (\mathbf{\nabla_j}\mathbf{\hat{  V}_{jk}} 
\mathbf{\nabla_j}\psi    ] \frac{1}{(m_j+m_k)c^2} (\Gamma_{jk})^{1/2}  \, d\xi, \, $  
\newline 
and the kinetic energy conversion rate, Eq. \ref{bp1},
\newline 
\indent \indent  \indent \indent  
$ (i\hbar/2m)[\psi^* \,\psi (\mathbf{\nabla_j}^2  \mathbf{\hat{V}_{jk}})     \; + \;   
2 (\mathbf{\nabla_j}\mathbf{\hat{V}_{jk}} \cdot  \mathbf{\nabla_j}\psi)  ]. \, 
$
\newline 
We set $ \, dt \; = \; t_E, \, $ and note that $ \, d\xi \; = \; \sqrt{dt}. \, $
With $ \, (\Gamma_{jk})^{1/2}  \, d\xi \, = \, 1, $ and dividing out common factors we get a ratio of  $  \, [ \hbar /(2 mc^2)] / \, t_E  \; = \; 1.69*10^{-5}. $ Since $t_E$ increases linearly with $r$ this is a maximum value (within the nonrelativistic limits). This ratio can be compared to that of the first-order relativistic deviations from the nonrelativistic formula for kinetic energy, $ p^2 / (2m). $

The first relativistic correction to the kinetic energy formula is 
$ - \frac{1}{8} p^4 / m^3c^2.  $ This is due to the increase of mass with speed. This can be rewritten as 
\newline 
$ - KE_{nr} * (\frac{1}{2})(KE_{nr} /mc^2) $ where $ KE_{nr} $ is the nonrelativistic expression for kinetic energy. If we assume that the kinetic energy is roughly within a factor of 2 of the potential energy, $ e^2/r$, then the correction is approximately $  (\frac{1}{4}) KE_{nr} * \alpha^2, $ where $\alpha^2$ is the square of the fine structure constant, $ 5.33*10^{-5}.$ So the deviation from the nonrelativistic formula is $ 1.33*10^{-5}.$ This is very similar to the stochastic discrepancy.

There are also several other forms of energy that nonrelativistic theory fails to account for. The most obvious of these is electromagnetic radiation. Although the deviation is a couple of orders of magnitude smaller than the one described above, the comparison is included here for the sake of completeness. We can use the Larmor radiation formula to estimate the classical ratio of radiated power to kinetic energy conversion. Kinetic energy is converted at the rate of 
$ \vec{\mathbf{F}} \cdot \vec{\mathbf{v}}. \, $ The Larmor formula is 
$ (\frac{2}{3}) * (e^2/c^3)a^2, $ where $a$ is the acceleration. With 
$ \vec{\mathbf{F}} \, = \, e^2/r^2, \, $  
$ \, v\, = \, \beta c, \,$ and $ a = \,  (e^2/m r^2), $ the comparison becomes 
$  (e^2/r^2) * \beta c \;  $  vs. 
$ (\frac{2}{3})  (e^2/c^3)(e^2/m r^2)^2. $\footnote{The acceleration is assumed to be entirely in the direction of motion; this minimizes the ratio.} 
So the ratio is $ (\frac{2}{3}) (1/\beta) [(e^2/r)^2 \, / (mc^2)^2]. $ If we set the nonrelativistic limit at $ \beta \, = \, \alpha, \, $ and $ \, r \, = \, r_B $ then this ratio becomes $\frac{2}{3}\alpha^3 \; = \; 2.59*10^{-7}.$

It is worth mentioning that the real part of the first-order stochastic term (which has been shown to integrate to zero) would have an approximately similar magnitude. Hence, the redistribution of kinetic energy that it implies relative to the standard theory would be very small.

\end{document}